\documentclass[aip, jcp, amsmath,amssymb,reprint]{revtex4-2}
\DeclareUnicodeCharacter{0301}{\'{e}}
\usepackage{graphicx}
\usepackage{media9}
\usepackage{hyperref}
\usepackage{xcolor}
\usepackage{color}
\usepackage{amsmath}
\usepackage{mathrsfs}
\usepackage{amsfonts}
\usepackage{amssymb}
\usepackage{varioref}
\usepackage{float}
\usepackage{dcolumn} 
\usepackage{bm}
\usepackage{relsize}
\usepackage{parskip}
\usepackage[flushleft]{threeparttable}
\newcommand{\cblue}{\color{black}}

\usepackage{epsfig}

\usepackage{bm}

\begin{document}
\preprint{APS/123-QED}

\title{Structural transitions of a Semi-Flexible Polyampholyte}

\author{Rakesh Palariya}
\email{rakesh20@iiserb.ac.in}

\author{ Sunil P. Singh  }
\email{spsingh@iiserb.ac.in}
\affiliation{Department of Physics,\\ Indian Institute Of Science Education and Research, \\Bhopal 462 066, Madhya Pradesh, India}

\date{\today}

\begin{abstract}
Polyampholytes (PA) are charged polymers composed of positively and negatively charged monomers along their backbone.  
The sequence of the charged monomers and the bending of the chain significantly influence the conformation and dynamical behavior of the PA.
Using coarse-grained molecular dynamics simulations, we comprehensively study the structural and dynamical properties of flexible and semi-flexible polyampholytes'. The simulation results demonstrate a flexible polyampholyte (PA) chain, displaying a transition from a coil to a globule in the parameter space of the charge sequence. Additionally, the behavior of the mean-square displacement (MSD), denoted as $<(\Delta r(t))^2>$, reveals distinct dynamics, specifically for the alternating and charge-segregated sequences.  The MSD follows a power-law behavior, where $<(\Delta r(t))^2> \sim t^\beta$, with $\beta \approx 3/5$ and $\beta \approx 1/2$ for the alternating sequence and charge-segregated sequence in the absence of hydrodynamic interactions, respectively. However, when hydrodynamic interactions are incorporated, the exponent $\beta$ shifts to approximately 3/5 for the charge-segregated sequence and 2/3 for the well-mixed alternating sequence. For a semi-flexible PA chain, varying the bending rigidity and electrostatic interaction strength ($\Gamma_e$) leads to distinct, fascinating conformational states, including globule, bundle, and torus-like conformations. We show that PA acquires circular and hairpin-like conformations in the intermediate bending regime. The transition between various conformations is identified in terms of the shape factor estimated from the ratios of eigenvalues of the gyration tensor. 
\end{abstract}

\maketitle


\section{\label{sec:level1}Introduction: }

Polyampholytes are a broader class of charged polymers comprising cationic and anionic residues on their backbone\cite{lowe2002synthesis,dobrynin2004polym,higgs1991theory};  with the random or periodic arrangements of charge sequence.\cite{silmore2021dynamics,dignon2020biomolecular,das2013conformations,long1998electrophoresis,rumyantsev2021sequence}
 The primary example of polyampholytes is proteins; typically, a protein consists of various amino acids; depending on the sequence of the amino acids, they can acquire well-defined three-dimensional folded or unfolded structures. The proteins that are unable to attain well-defined folded structures are known as intrinsically disordered proteins (IDPs). Many biopolymers and IDPs can be classified as polyampholytes. The IDPs play a crucial role in many biological processes, despite having unfolded structures, specifically in regulating translation\cite{dyson2005intrinsically}, transcription, inter-cellular signaling\cite{tantos2012intrinsic}, controlled spatial and temporal signaling\cite{wright2015intrinsically}, cell division, aggregation and segregation of microtubules, potential role in various protein-related diseases\cite{dima2004proteins}, in assisting liquid-liquid phase separation (LLPS) in membrane-less organelles\cite{shin2017liquid,shin2017liquid,banani2017biomolecular,brangwynne2009germline}, etc.

 The amino acid sequence and weak hydrophobicity primarily dictate the unfolded protein structures in physiological conditions. Weak hydrophobicity and relatively higher net charge are essential factors that make it more likely for IDPs to acquire coil-like structures in physiological conditions.
The electrostatic attraction among positive and negative side groups dominate over the usual solvent-mediated hydrophobic interactions\cite{bright2001predicting,dobrynin2004polym,jiang2006phase,castelnovo2002phase,shusharina2005scaling,wang2006regimes,higgs1991theory,cheong2005phase,everaers1997complexation,barbosa1996phase,mao2010net,uversky2002does}; therefore, such proteins lack the ordered folded structures\cite{xie2007functional,tompa2012intrinsically} and acquire either good-solvent or theta-solvent like conformations in the aqueous medium. Thus, structural behavior such as the radius of gyration follows the scaling relation, $R_g\sim N^{\nu}$, with power-law exponents $\nu=3/5$ or $1/2$, respectively.\cite{samanta2018charge}

The specificity of the charge sequence is another key parameter in dictating the structural behavior of polyampholytes, in addition to other physical parameters such as total fractional charge, the dielectric constant of the medium, salt concentration, rigidity of chain, etc. \cite {das2013conformations,samanta2018charge,bianchi2020relevance,danielsen2019molecular} The variation in the sequence of the positive and negative charges can lead to a large-scale conformational change despite having the same total number of charges on the backbone.\cite{das2013conformations,danielsen2019molecular,samanta2018charge,danielsen2019molecular,muller2010charge,sundaravadivelu2024sequence}   For example, conductivity, relaxation, and solution properties such as the precipitation and micelle formation can be significantly influenced by charge distribution.\cite{gohy2000aggregates,silmore2021dynamics,devarajan2022effect,goloub1999association,radhakrishnan2021collapse}

A substantial amount of work has been devoted to unraveling the structural behavior of block polyampholytes by varying block lengths and salt concentration in the theory and molecular simulations.\cite{danielsen2019molecular,dobrynin2004polym,samanta2018charge,jiang2006phase,castelnovo2002phase,shusharina2005scaling,wang2006regimes,higgs1991theory,cheong2005phase,everaers1997complexation,barbosa1996phase,imbert1999conformational} Long-range attractive interactions among various blocks of the chain drives into the globular phase in a salt-free solution.\cite{edwards1980phase,kantor1991polymers,dobrynin1995flory}  In addition, it has been shown that the overall charge-imbalanced polyampholytes may collapse into a globular phase, and stronger charged PA can acquire pearl-necklace-like conformations where globular structures are connected by stretched chains.\cite{dobrynin2004polym,dobrynin1995flory,jeon2005molecular,long1998electrophoresis} 
Scaling theory has been mainly focused on the flexible diblock polyampholytes, which exhibit the shift from the globular state to the tad-polar structure for symmetric and asymmetric block PAs, respectively.\cite{wang2006regimes,shusharina2005scaling,castelnovo2002phase}
Few recent works accounting for a large number of specific sequences and the role of total fractional charges have explored conformational behavior via atomistic simulations, which disclose the monotonic decrease of the radius of gyration with the sequence charge decoration ($SCD$) parameter.\cite{das2013conformations,sawle2015theoretical} More importantly, this study demonstrates the radius of gyration is a nearly universal function $SCD$ despite having very different charge sequences at the same charge composition.\cite{das2013conformations} 

Previous studies have extensively focused on the structural behavior of flexible polyampholytes, considering a large number of specific charge arrangements and salt concentrations\cite{sawle2015theoretical,das2013conformations}; however, less has been discussed about the role of semi-flexibility.\cite{baumketner2001helix,baratlo2010brushes,akinchina2007diblock}
It is noteworthy that change in the chain's bending rigidity for a neutral polymer can cause large-scale structural changes, from globule, torus, rod-like, and ordered states to topologically different knotted states.\cite{zierenberg2016dilute,de1979scaling,seaton2013flexible,marenz2016knots,huang2016simple,aierken2023impact,gordievskaya2018interplay}

The phase behavior of the grafted diblock semiflexible spherical PA brushes presents many interesting structures, such as star-like structures, bundle-like and ordered inner cores, and disoriented outer cores.\cite{akinchina2007diblock}
Despite the ample relevance of polyampholytes in bio-polymers, the effect of rigidity on their structure and dynamics has been vastly ignored.

\begin{figure*}
\includegraphics[width=\textwidth]{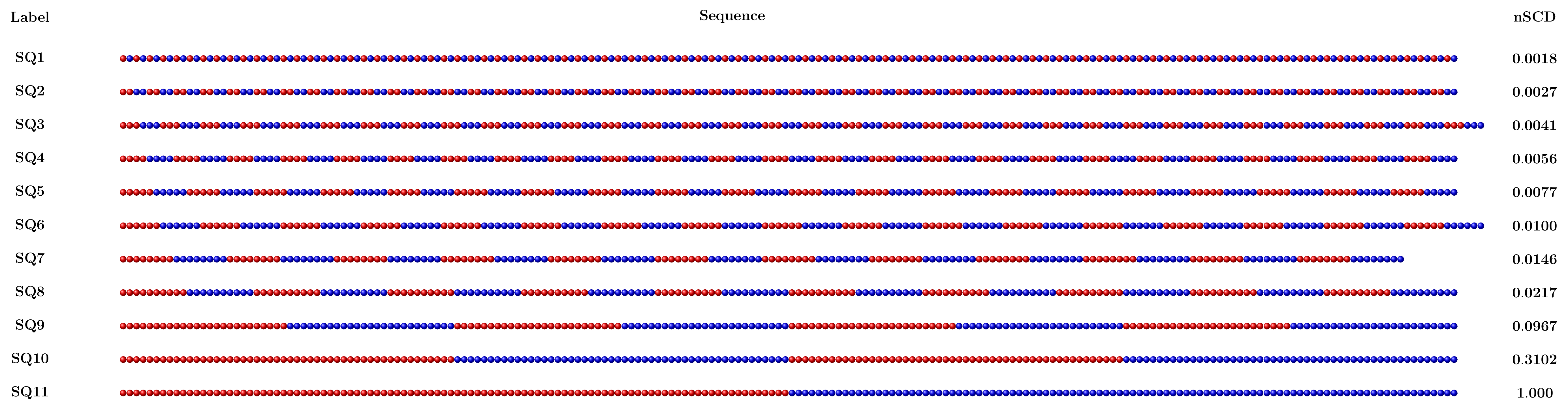}
\caption{Sequences used for the simulation. Column 1 shows the label of sequences; column 2 shows the sequence in which the red monomer is positively charged and the blue monomer is negatively charged; and column 3 shows the normalized sequence charge decoration (nSCD).}
\label{Fig:sequence}
\end{figure*} 

 Considering these aspects, our current work considers a coarse-grained model of polyampholytes composed of periodically arranged positively and negatively charged blocks, with a periodicity of $n_B$ monomers of length $L_B$, of equal numbers in a salt-free solution. The number of blocks varies while the chain length is fixed to account for various charge sequences, thus resulting in the variation of the $SCD$. The definition of the charge segregation parameter can be referred to in the simulation model. The sequence of PA and variation of block lengths ($L_B$) are systematically shown in Fig.~\ref{Fig:sequence}.
The simulation results can be divided into two major sections: one considers the role of the charge sequence on the structural and dynamical behavior of a flexible PA chain, and the second part presents the role of semi-flexibility in its structure.

 Our simulation results provide valuable insights into the structural behavior of flexible polyampholyte (PA) chains, particularly the impact of varying sequences of charged monomers. We observe a continuous transition from a coil-like state to a $\theta$-solvent state and eventually to a globular state as the block length of the PA chain increases. This transition is revealed in the parameter space of the charge sequence, with notable characteristics observed near the $\theta$-point at which the PA chain exhibits ideal polymer-like behavior.  Our results elucidate that the radius of gyration, when normalized by the value at $\theta$-point, aligns closely across various electrostatic strengths, underscoring the robustness of this behavior irrespective of the electrostatic strengths. Further, we delve into the dynamic behavior of the monomers within the PA chain by analyzing their mean-square displacement (MSD), with and without hydrodynamic interactions. We gain insights into how the sequence of charged monomers and hydrodynamics affect the internal dynamics of the PA chain. Our results illustrate in the charge segregated state monomers exhibit sub-diffusive behavior their MSD displays  $<(\Delta r(t))^2> \sim t^\beta$ in the intermediate time regime, with an exponent $\beta \approx 1/2$ in the absence of the hydrodynamic interactions but in the presence of hydrodynamic interactions, the exponent shifts to approximately $\beta\approx 3/5$.

The second part presents the role of semi-flexibility on the conformational behavior of the PA chain; we observe a large-scale structural transition characterized by a shift from elongated globular states to circular or torus states. To discern between these conformations, we calculate the shape factor and analyze the energies of associated states.  Additionally, we present a detailed phase diagram of the diblock PA chain, mapping out its conformations in bending rigidity and electrostatic interaction strength $\Gamma_e$, offering a comprehensive overview of its conformational landscape under such conditions. 

This paper is organized as follows: the simulation model is introduced in Section II, and the results of the flexible and semiflexible PA are presented in Sections III and IV, respectively. All the results are summarized in Section IV.

\section{Model}
We model a linear polyampholyte chain comprised of an equal number of positive and negative-charged monomers such that the net charge of the PA chain is zero in a salt-free solution. The sequence of the charge pattern along the backbone is considered periodic in blocks.   The block length, equivalently referred to as the number of charged monomers in a block, is considered to be an independent parameter that is varied from the diblock (the largest block) to the alternating sequence of positive and negative charges (the smallest block) for a given polymer length ($N$).  

The interaction potentials in our simulation model include excluded volume ($U_{LJ}$), long-range electrostatic interaction ($U_C$), connectivity of the monomers ($U_F$), and bending rigidity ($U_B$), which makes the total interaction potential $U = U_F + U_{LJ} + U_C+ U_B$. The linear connectivity of neighboring monomers is achieved via finite extensible nonlinear elastic (FENE) potential\cite{hsieh2006brownian, LAMMPS} 
\begin{equation}
 U_{F}\left(r\right) = -\sum_{i=1}^{N-1}\frac{1}{2}\kappa_{F}R^{2}_{0}\ln\left(1 - \frac{r_{i,i+1}^2}{R_{0}^{2}}\right) 
 \label{Eq:u_f}
\end{equation}
where $\kappa_{F}$ is the spring constant, $R_0$ dictates terms of the maximum extension of bond,  ${\bm r}_i$ is the position vector of the $i$th monomer, and ${\bm r}_{i,i+1}={\bm r}_{i+1}-{\bm r}_i$ is the $i^{th}$ bond vector of the PA chain.
The excluded volume interaction among the monomers, which precludes the overlap of the monomers, is employed by the truncated and shifted Lennard-Jones potential (LJ) given by
\begin{eqnarray}
 U_{LJ} =\sum_{i>j}^{N}  4 \epsilon_{LJ} \left[ \left(\frac{ \sigma}{r_{i,j}}\right)^{12} - 
 \left(\frac{\sigma}{r_{i,j}}\right)^{6} + \frac{1}{4} \right].
 \label{Eq:u_lj}
\end{eqnarray}
The above potential, $U_{LJ}$, is applicable for $r_{i,j} \leq 2^{1/6}\sigma$; otherwise, $U_{LJ}=0$.
Further, $r_{ij}$ is the distance between two monomers, $\epsilon_{LJ}$ is the interaction parameter, and $\sigma$ is the LJ diameter.

The semiflexibility is introduced by the cosine bending potential\cite{Rubi:2003}, $U_{B} (\theta) =\sum_{i=1}^{N-2} {\kappa_b}(1 - \cos(\theta_{i,i+1}))$, 
where $\theta_{i,i+1}$ is the angle between adjoining bonds and $\kappa_b$ denotes the bending stiffness. In the case of the neutral chain,  $\kappa_b$ directly relates to the persistence length ${\ell}_p$ of the chain, where  ${\ell}_p$ is the length over which a chain's tangent-tangent correlations decay to $e^{-1}$ of its value.\cite{doi:86,Muthu_Book_2011,Gennes_SCP_1979,Rubi:2003}
For the PA chain, the persistence length may depend on various other factors, such as the sequence of charge, bending rigidity ($\kappa_b$), and strength of the electrostatic interactions.
The long-range Coulomb potential imposes the electrostatic interaction among the charged monomers 
\begin{equation}
U_C= {\frac{1}{4\pi\varepsilon} } 
\sum_{\mathbf{n}}{}^{'} \sum_{i=1}^{N} {\sum_{j=1}^{N}} \frac{z_iz_j e^2}{|\mathbf{ r}_{i,j}+\mathbf{n}L|}. 
\label{Eq:u_e}
\end{equation}

Where $z_{i} = \pm e$ is the valency of the monomers and $L$ is the size of the primary periodic box. $\varepsilon$ is the dielectric constant of the solvent for the simplicity of the model, it is assumed to be homogeneous throughout the medium. The first summation is over all the periodic images of the primary box excluding the primary box, i.e., ${\bm n}(n_x,n_y,n_z)=(0,0,0)$, which excludes the self-interaction.  To compute the long-range force, the particle-particle/particle-mesh (PPPM) technique is employed, which is computationally efficient over Ewald summation for a relatively bigger system.\cite{deserno1998mesh_1,deserno1998mesh_2,LAMMPS}
The strength of the electrostatic interaction ($\Gamma_e$) among the charge monomers can be expressed in terms of the Bjerrum length  ${\ell}_B = \frac{e^2}{4\pi \epsilon k_BT}$ and LJ diameter   $\sigma$, given as $\Gamma_e={\ell}_B/\sigma$.  Note that the ${\ell}_B$ is the distance at which the electrostatic energy between two charged monomers equals the thermal energy. For various biological polymers like proteins, RNA, DNA, etc., $\Gamma_e$ spans the range of $2-4$.\cite{Muthu_Book_2011,Rubi:2003}

We perform Langevin dynamics simulations of a single PA chain in a cubic box with periodic boundary conditions (PBC). The equation of motion of the  PA  monomer is written as 
\begin{equation}
    m_i \frac{d^2{\bf r}_i}{dt^2} = - \nabla_i U - \zeta {\bf v}_i  + {\bf F}_i^r(t), 
    \label{Eq:m_i}
\end{equation}
where $m_i$, $r_i$, and $v_i$ are the monomer's mass, position, and velocity, respectively.  The first term on the RHS of the Eq.~\ref{Eq:m_i} corresponds to force due to the conservative potentials (U), the second term -$\zeta {\bf v}_i$ is the viscous drag where $\zeta$ is the drag coefficient, and the third term ${\bf F}_i^r$ is the thermal noise imparted by the implicit solvent. The thermal noise  is a normal distribution along with its first moment  $<F_i^r>=0$, and the second moment obeys the  fluctuation-dissipation theorem:
$    <{\bf F}_i^r(t) \cdot {\bf F}_j^r(t')> = 6 \zeta k_BT\delta(t-t')\delta_{ij},$
where $k_BT$ is the thermal energy.\cite{allen87}

The sequence charge decoration  parameter (SCD)\cite{sawle2015theoretical,huihui2018modulating} is computed as
\begin{equation}
    SCD = \frac{1}{N} \sum_{i=2}^{N} \sum_{j=1}^{i-1} z_i z_j \sqrt{i-j},
    \label{Eq:SCD}
\end{equation}
where $z_i$ and $z_j$ are charges on the $i^{th}$ and $j^{th}$ monomers, respectively.
The SCD parameter measures the spread of positive and negative charges on the PA. For example, in the case of the diblock PA, the first half of the chain is positively charged, and the latter half is negatively charged; thus, the value of $SCD$ will be maximum; see Fig.~\ref{Fig:sequence}-a. Similarly, SCD decreases for the smaller blocks in PA. From the perspective of comparison of different chain lengths, the SCD can be normalized so that it is $1$ for the fully segregated chain, i.e., for the diblock sequence, and a very small value for the alternating blocks or homogeneously mixed sequence.

{\it Simulation Parameters:} All the physical parameters are presented in dimensionless form using specific simulation units.  {\cblue The lengths is presented in units of the LJ diameter $\sigma$, energy in units of $\epsilon_{LJ}$, and time in units of $ \tau = \sqrt{\frac{m \sigma^2}{k_BT}}$.
The LJ diameter is set to be $l_0/\sigma \approx 0.98$ for the chosen parameter, thermal energy ${\epsilon_{LJ}}/{k_BT} = 1$, the FENE spring constant $\kappa_F=30$ and $R_0=1.5$, and bending parameter $\kappa_b$ in units of ${\epsilon_{LJ}}/{\sigma^2}$ is varied in the range of 0 to 30.} The chain length is $L_c/\sigma =200$, corresponding to the number of monomers $N=200$; however, to accommodate certain block lengths for the nearly same chain length, the number of monomers is considered slightly different than $200$. For example, for the case of block lengths 8, 16, and 32, the chain lengths are taken $192,204$, and $196$, respectively; see Fig.\ref{Fig:sequence}. The solution's dielectric constant $\varepsilon$ is considered uniform, and the electrostatic interaction strength ($\Gamma_e$) varies in the range of $0.5-6$.  In addition, most of the simulation results for the semi-flexible PA are presented for the chain length $N=64$. The particle-particle particle-mesh (PPPM) technique is employed here to compute the long-range Coulomb interactions. 
The parameters for the PPPM method are chosen to achieve a numerical error in the force calculations of the order of $10^{-4}$.\cite{deserno1998mesh_1,deserno1998mesh_2}  At dilute concentrations, the effect of the counterions on the PA's structure is negligible; therefore, they are not considered for simplicity.

We solve the Langevin equation using the velocity-Verlet algorithm for a time step of $ \delta t = 5\times 10^{-3}\tau$, with the help of the molecular dynamics simulation package LAMMPS.\cite{LAMMPS}
All the simulations are performed for a total simulation time of ($2 \times 10^5$) time units, and out of that, ($5 \times 10^4$) time units are used for the equilibration of the system. Each presented data point is averaged over at least twenty independent simulations for better statistical accuracy. 

\section{Flexible Polyampholyte}
A flexible PA chain undergoes a large conformational change, specifically coil-to-globule transition, on the variation of charge sequence. \cite{das2013conformations,sawle2015theoretical,danielsen2019molecular,srivastava1996sequence,dinic2021polyampholyte}
In particular, a random sequence and homogeneously mixed sequence exhibit a random coiled state, while a charge-segregated PA sequence attains a compact globular state.
In the subsequent subsection, we emphasize the role of electrostatic interaction strength ($\Gamma_e$) on the structure of a flexible PA chain. Further, we compute the static structure factor to determine the scaling behavior of the chain and $\theta$-point.\cite{Rubi:2003,doi:86,Gennes_SCP_1979,Muthu_Book_2011} Additionally,  we also investigate the internal dynamics of the flexible chain by probing the MSD of a monomer in the presence of long-range hydrodynamic interactions (HI).\cite{ Gompper_APS_2009,Kapral_ACP_2008,Singh_2014_JCP} 

\begin{figure}[tb]
\includegraphics[width=\linewidth]{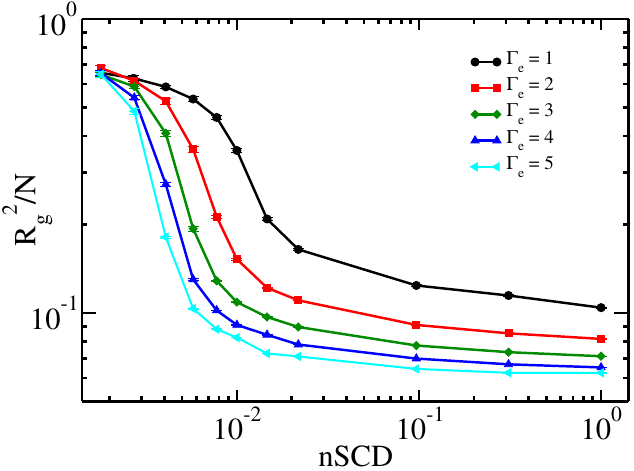}
\caption{The mean-square radius of gyration ($R_g^2/N$) of a flexible PA chain ($\kappa_b=0$) as a function of normalized sequence charge parameter (nSCD) for various $\Gamma_e$ as displayed in the plot.}
\label{Fig:rg_at_0}
\end{figure}

\subsection{ Radius of Gyration}

 The radius of gyration $R_g$ characterizes the average size of the chain, which can be computed as, 
\begin{equation}
    R_g^2 = \frac{1}{N} \Biggl\langle \sum_{i=1}^{N} \left({\bm r}_i - {\bm R}_{cm} \right) ^2 \Biggr\rangle,
    \label{Eq:R_g}
\end{equation}
where $R_{cm}$ is the center of mass of the PA chain. The computed values of the $R_g^2$ for various $\Gamma_e$ is plotted in 
Fig.~\ref{Fig:rg_at_0} as a function of the normalized sequence charge parameter $nSCD$, normalized by SCD of the diblock, i.e., the charge segregated conformation.  
The plot shows a monotonic variation of the radius of gyration, where it sharply reaches a plateau value for large $nSCD$s that correspond to larger regular blocks.
The sharp decrease of $R_g^2$ in a narrow window of $nSCD$ indicates that the PA chain undergoes a continuous transition from open coil-like states to compact globular-like states for a larger block length. The transition becomes more prominent and shifts towards smaller $nSCD$  values for larger $\Gamma_e$ as indicated in  Fig.\ref{Fig:rg_at_0} for various $\Gamma_e$.

For the smaller blocks, the electrostatic interaction is screened locally; thereby, the PA attains a random coil-like state. Upon increasing block lengths $L_B$, electrostatic attraction among opposite charges grows, forming a complex of oppositely charged blocks, thereby shrinking the chain. As  $nSCD\rightarrow 1$, $R_g$ decreases significantly, and the chain acquires a compact globular structure. The PA chain displays qualitatively similar behavior for higher electrostatic interaction strength; however, the transition shifts towards smaller values of $SCD$, and the chain acquires a relatively compact structure for larger values of $nSCD$, due to strong electrostatic attraction among oppositely charged blocks.

 \begin{figure}[tb]
\includegraphics[width=\columnwidth]{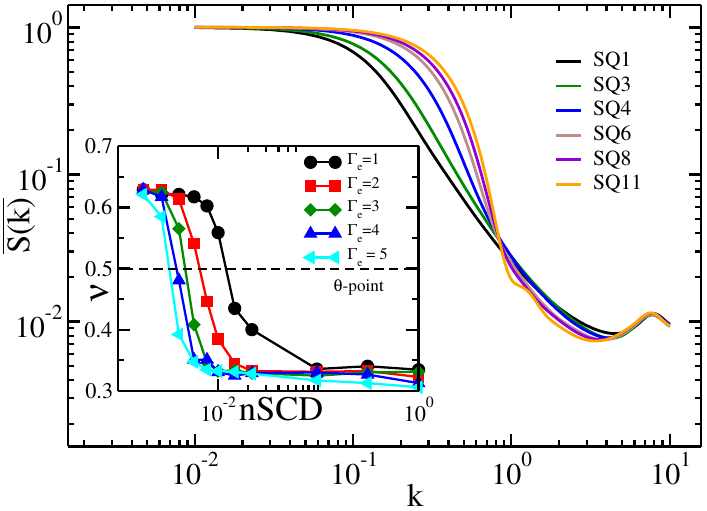}
\caption{The normalized static structure factor $\overline{S(k)}=S(k)/S(0)$) of a flexible PA as a function of the magnitude of wave vector $k$ at $\Gamma_e= 3.0$ for various sequences indicated in the plot. The variation of the Flory exponent ($\nu$),  fetched by fitting the power-law in $\overline{S(k)}$, as function nSCD for various  $\Gamma_e$. The dashed line marks $\theta$-point, i.e., $\nu\approx 1/2$ on the plot.}
\label{Fig:s_factor}
\end{figure}

\subsection{ Static Structure Factor} 
We compute the static structure factor of a single PA chain to identify the coil-to-globule transition point while varying the charge sequence. 
The structure factor $S(k)$ describes the correlation of the density fluctuations at different length scales, which can be measured directly in the scattering experiments.\cite{Rubi:2003,doi:86,de1979scaling} The statistics of the PA chain and its scaling exponent ($\nu$) can be estimated in the intermediate regime of the $k$-vector. We can compute $S(k)$ numerically from
\begin{equation}
   S(k) = \frac{1}{N} \Bigg\langle \sum_{j=1}^{N} \sum _{k=1}^{N} e^{-i {\bf k}\cdot({\bm r}_j - {\bm r}_k)} \Bigg \rangle ,
    \label{Eq:s_k}
\end{equation}
where ${\bm k}$ is the scattering wave vector. The computed structure factor $S(k)$ for various $nSCD$ parameters is presented in Fig.~\ref{Fig:s_factor} at $\Gamma_e=1.0$. In equilibrium, $S(k)$ is isotropic and depends on the magnitude of the ${\bm k}$-vector. As expected, the $S(k)$ has a plateau region for very small wave vectors ($R_g^{-1} << k $) and a nearly universal curve for large vectors, indicating that at the scale of the monomer, all the chains have the same structure irrespective of their sequence. The plateau regime of $S(k)$ for larger block lengths is larger ($SQ11$) and shrinks further for smaller block lengths ($SQ1$). The broader plateau regime signifies a more compact state (i.e., small $R_g^2/N$).
{\cblue Notably, the appearance of a weak plateau regime for the diblock PA chain  ($SQ11$) in the range of $0.7\leq k\leq2$ is to be due to the globular structure of the PA chain at this parameter\cite{mukherji2017depleted}. }

In the intermediate regime ($R_g^{-1} << k << l_0 ^{-1}$), the $S(k)$ falls with a power law given by $ S(k) \propto k^{-\frac{1}{\nu}} $, where the   $\nu$ is the polymer's scaling exponent.\cite{Rubi:2003, doi:86,Gennes_SCP_1979,Muthu_Book_2011,flory1953principles,birshtein1987theory} For the clarity of the plots, $S(k)$ is shown only for a few charge sequences (nSCD). From Fig.~\ref{Fig:s_factor}, we can estimate Flory's exponent $\nu$ by employing a power law fit given as, $ S(k) \propto k^{-\frac{1}{\nu}} $. The inset of Fig.~\ref{Fig:s_factor} displays the obtained Flory exponent for all the parameters.

\subsection{ Coil-globule Transition} 
The exponents $\nu $ clearly indicate the continuous transition from the random coil-like chain in good solvent $\nu=3/5$ (small block lengths) to ideal chain conformations ($\theta$-point $\nu=1/2$) followed by a globular-like structure for larger blocks with $\nu\approx1/3$, as the block length is varied, see Fig.~\ref{Fig:s_factor}.  The dashed line illustrates the obtained $\theta$ point, which shifts towards smaller values of $nSCD$ (i.e., smaller blocks). This indicates that for larger electrostatic strengths $\Gamma_e$, PA acquires the ideal chain behavior even for smaller block lengths.\cite{flory1953principles,birshtein1987theory} 
The shift of $\theta$-point is a consequence of the stronger electrostatic attraction for large $\Gamma_e$; this subsides the swelling of the PA due to excluded volume repulsive interactions. The transition from a collapsed conformation to an extended conformation results from the cancellation of the attractive electrostatic and repulsive excluded volume interactions.

 \begin{figure}[t]
\includegraphics[width=\linewidth]{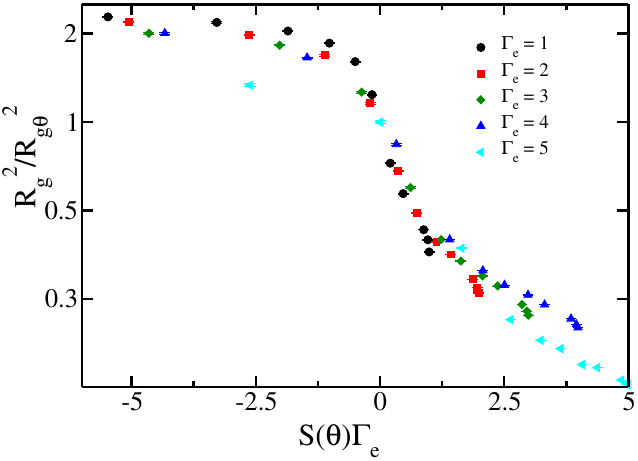}
\caption{ The normalized radius of gyration  $R_g^2/R_{g\theta}^2$ as function of  $S(\theta)\Gamma_e = (1 - nSCD_\theta/nSCD)\Gamma_e$,  where $R_{g\theta}^2$ is the value of the radius of gyration at $\theta$-point. }
\label{Fig:normalized}
\end{figure}

Now we present the radius of gyration normalized by $R_{g\theta}^2$ at $\theta$-point obtained from Fig.~\ref{Fig:rg_at_0}, where $\theta$-point is obtained from Fig.~\ref{Fig:s_factor} employing linear interpolation expression. $R_g^2/R_{g\theta}^2$ is presented as a function of $S(\theta)= (nSCD-nSCD_{\theta})/nSCD$, as in the case of coil-globule transition, the expansion factor is presented with the interaction parameter. \cite{Rubi:2003} Interestingly,  $R_g^2/R_{g\theta}^2$  superimposes irrespective of $\Gamma_e$, if  $S(\theta)= (nSCD-nSCD_{\theta})/nSCD$ is linearly scaled with the $\Gamma_e$.  This suggests the conformational properties of the PA are essentially identical if the $\theta$-point is scaled with $\Gamma_e$; see Fig.~\ref{Fig:normalized}. 

Note that here, coil-globule transition is driven by varying the block length of charge sequence that tunes the attractive interactions, unlike the neutral polymer, where the temperature is varied to achieve the transition. The parameter $\chi$ describing the interaction's strength is a function of the block length. Accordingly, using the analogy of the neutral polymer near the $\theta$-point $z\approx \frac{(T-T_\theta)}{T} N^{1/2}$, where $z=R^2_g/R_{g\theta}^2$ is the expansion factor, gives $g_T\approx \left(\frac{T}{(T-T_\theta)}\right)^2$\cite{Rubi:2003,Muthu_Book_2011,Gennes_SCP_1979}, where $g_T$ is the number of thermal blobs. This can be estimated for the PA, as Fig.~\ref{Fig:normalized} indicates the $g_T \approx \left(\frac{SCD}{(SCD-SCD_\theta)\Gamma_e} \right)^2$, thus the thermal blobs is reduced by the larger electrostatic strength, as expected, $g_T \approx \Gamma_e^{-2}$.  In summary, the transition from coil to globule exhibits a similar characteristic in the parameter space of the charge sequence near the $\theta$-point.

\subsection{ Dynamics}
Now, we present the internal dynamics of the flexible PA chain; for this purpose, we also incorporate solvent-mediated long-range hydrodynamic interactions. 
 For this, we employ multi-particle collision dynamics simulations (MPC), which is an explicit solvent-based simulation technique; the details of the method and its coupling with the polymer can be found in the references cited herein.\cite{Gompper_APS_2009,Singh_JPCM_2012,Kapral_ACP_2008} To compare the dynamics at the same parameters and solvent viscosity, the simulations without hydrodynamics are also performed within the framework of the MPC  at the same solvent viscosity.

The internal dynamics of the PA can be quantified from the MSD of the chain's monomer. The MSD quantifies the local relaxation of the monomers at a short time and the diffusive behavior of the chain in the longer-time regime.  The MSD  of a monomer  in the center of the mass frame can be computed as, 
\begin{equation}
   \left<(\Delta r(t))^2 \right> = \left< ({\bm r}(t)-{\bf R}_{cm}(0))^2 \right>,
\end{equation}
where the angular bracket represents the ensemble average. In addition, the MSD is also averaged over all the monomers of the PA chain. In the short time interval  $t< k_BT/\gamma$, the monomers move ballistically; therefore, $ <(\Delta r(t))^2> \sim t^2 $. In the intermediate regime, the connectivity with the chain and other interactions dictates the relaxation and dynamics of the monomers. Thereby, a sub-diffusive regime appears due to constriction, where MSD of the monomer can be expressed as $ <(\Delta r(t))^2> \sim t^\beta$, the power-law exponent for the case of the ideal chain is  $\beta=1/2$.\cite{Muthu_Book_2011,doi:86,Rubi:2003}

\begin{figure}
\includegraphics[width=\columnwidth]{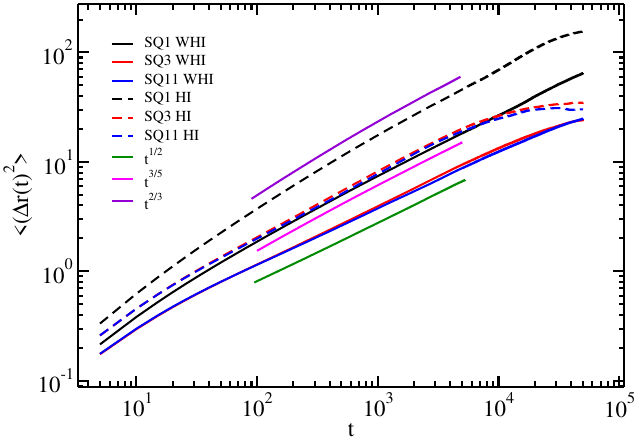}
\caption{The MSD (  $ <(\Delta r(t))^2>$) of a PA monomer w.r.t. center-of-mass of the chain for various block lengths 1, 3, and 100. The dashed lines display the MSD of a  PA monomer with hydrodynamics (HI) at the same solvent viscosity $\eta_s=8.7$. The solid line displays the MSD at the same parameters without hydrodynamics (WHI). In the intermediate regime,  the subdiffusive behavior is displayed by half-solid lines with their exponents mentioned in the plot.   } 
\label{Fig:msd}
\end{figure}

Figure~\ref{Fig:msd} displays the computed values of the MSD of the PA monomer for the three different values of the block lengths. The PA chain acquires the ideal and globule-like structures for the block lengths $3$  and 100 (diblock), respectively.  Figure~\ref{Fig:msd} displays that the  MSD for these parameters is nearly identical. The MSD in the intermediate regime displays $<(\Delta r(t))^2> \sim t^\beta$ with  $\beta\approx1/2$ for both chains. This confirms that the internal dynamics in this regime are similar to those of the ideal polymer. For the case of the well-mixed charge sequences, the MSD in the sub-diffusive regime is characterized by  $<(\Delta r(t))^2> \sim t^\beta;$ with the power-law exponent $\beta\approx 3/5$, in the absence of hydrodynamic interactions.

In the presence of the hydrodynamic interactions, the dynamics of the PA chain and its monomers are relatively faster, as captured in the MSD; besides this, the sub-diffusive exponents are also larger than that.  For a well-mixed sequence, it goes up to $\beta\approx 2/3$, but for $\theta$-solvent and charge-segregated states, $\beta\approx 3/5$, which is higher than the value obtained without hydrodynamic interactions.  This suggests that the monomer dynamics are relatively faster when long-range hydrodynamic interactions are included. In addition, the dynamics of the monomers are relatively faster in the case of open structures for alternating blocks. Notably, the dynamics get slower for the larger block lengths.

 \section{Semi-flexible Polyampholyte}

\begin{figure}
\includegraphics[width=\columnwidth]{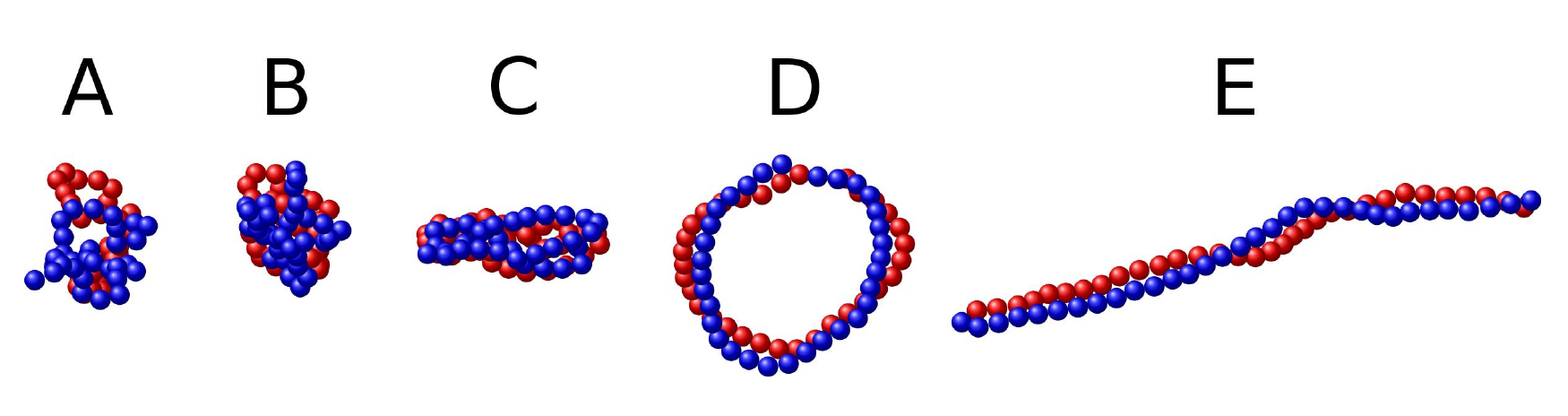}
\caption{ Various conformational states of a diblock PA in the parameter space of $\kappa_b$ and $\Gamma_e$ for $N=64$. Coil-state (A), Globular-state (B), elongated globule (C), circular state (D), and hairpin (E)  states.  }
\label{Fig:diblock_conformation}
\end{figure}

For a semi-flexible  PA chain, the interplay of the bending and electrostatic interactions offers various intriguing structures, which are absent in the flexible limit.
The role of bending $\kappa_b$, the strength of electrostatic interactions $\Gamma_e$, along with the charge sequence length ($L_B$), on the structure of the PA is characterized. The emergent structures of the PA chain for charge sequence block length $L_B=32$ (diblock) are depicted in Fig.~\ref{Fig:diblock_conformation}. Similarly, we have shown various structures emerging on the variation of bending parameter (progressing from left to right) in SI-Fig.1 for the diblock and tetra block PA.   In the rest of the discussion, we limit our consideration to a smaller chain length of $N=64$  to avoid the complexity of meta-stable states that emerge for the longer chains.

{\cblue To avoid trapping in the metastable conformations of a semi-flexible PA chain, we use sequential annealing of the temperature from a high-temperature regime to step down gradually to the desired low temperature. The molecular dynamics simulations are started at $k_BT=5$ with starting as a rod-like conformation.  We slowly anneal the temperature sequentially in steps of $0.25k_BT$ until the system reaches the target temperature ($k_BT=1$).  For each temperature, the system is equilibrated to $2 \times 10^4$ simulation time steps before it is further lowered. The final simulation run is performed $5 \times 10^4$ simulation steps. By sequentially lowering the temperature in small steps, we allowed the PA sufficient time to explore various conformational states. 
We executed extensive simulations to enhance statistical accuracy for these intricate states and discern stable conformations from various metastable states for presented charge sequences. Each dataset comprises 60 independent simulation runs for the semi-flexible case to acquire good statistics.}\\


\subsection{Diblock Chain} 

In this section, we discuss the conformational transitions of the diblock PA by varying $\Gamma_e$ and the chain's bending rigidity ($\kappa_b$), see Fig.~\ref{Fig:diblock_conformation}.
The diblock PA chain acquires coil-like conformations for  $\Gamma_e<2$ and $\kappa_b<5$. However, as we progressively increase $\Gamma_e$ and $\kappa_b$, we observe a transition from globular-like (see SI-Movie-1) form to folded bundles  (see SI-Movie-2) followed by circular states  (see SI-Movie-3). To quantify these states, we compute the radius of gyration, shape factor, and the average distance between various segments in the following section of the manuscript.

\begin{figure}[th]
\includegraphics[width=\columnwidth]{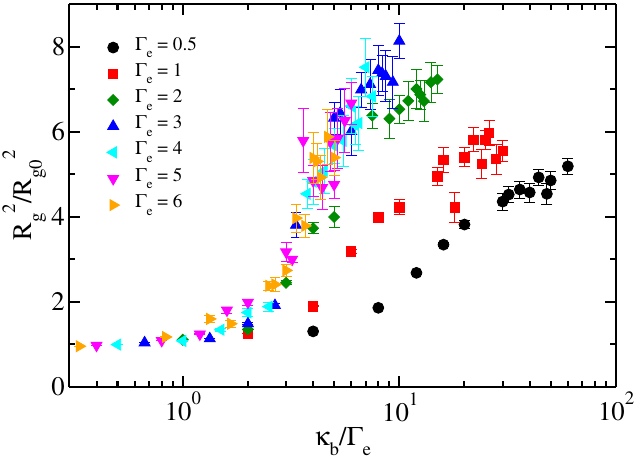}
\caption{ The normalized mean-square radius of gyration $R_g^2/R_{g0}^2$ of the diblock PA chain as a function of the bending rigidity $\kappa_b$ scaled with the electrostatic interaction strength $\Gamma_e$ for various $\Gamma_e$ as shown in the figure.  Here, $R_{g0}^2$ is the mean-square radius of gyration of the chain at $\kappa_b=0$.  }
\label{Fig:radius_vs_lb}
\end{figure}

\subsubsection{Radius of Gyration} 
The average size of a semi-flexible polyampholyte (PA) chain can be effectively characterized by examining the mean-square radius of gyration ($R_g^2$). Figure~\ref{Fig:radius_vs_lb} presents the normalized mean-square radius of gyration $R_g^2/R_{g0}^2$ as a function of the bending rigidity $\kappa_b/\Gamma_e$ for a diblock PA chain across various electrostatic strengths ($\Gamma_e$).  The $R_g^2/R_{g0}^2$ remains the same for $\kappa_b/\Gamma_e<1$.  As $\kappa_b/\Gamma_e>1$ is incremented, bending increases; hence, the chain monotonically swells for all $\Gamma_e$. In the latter case, bending dominates over the electrostatic attraction, while in the former case, attraction overpowers the bending; therefore, the chain acquires a coil or globule shape. 
The PA chain predominantly assumes a coil-like conformation for small $\Gamma_e \leq 1$, however, for larger $\kappa_b$, a smooth transition is observed from the coil to a hairpin-like conformation, as Fig.~\ref{Fig:radius_vs_lb} reflects for $\Gamma_e=0.5$ and $1.0$.

Due to stronger electrostatic attraction among charge sequences, the PA chain acquires a compact globular state for $\kappa_b/\Gamma_e< 1$. Upon an increase in bending, the chain starts expanding as $R_g^2/R_{g0}^2$ increases,  beyond $\kappa_b/\Gamma_e>1$.  More importantly,  Fig.\ref{Fig:radius_vs_lb} indicates that the $R_g^2/R_{g0}^2$ is nearly independent of electrostatic interaction strength, where all the curves superimpose on top of each other with $\kappa_b/\Gamma_e$.
 A narrow plateau indicates the elongated shape globule for $\Gamma_e=4,5$ and $6$,  In the $1<\kappa_b/\Gamma_e$ $<3$ regime.

\begin{figure}[t]
{\includegraphics[width=0.48\columnwidth]{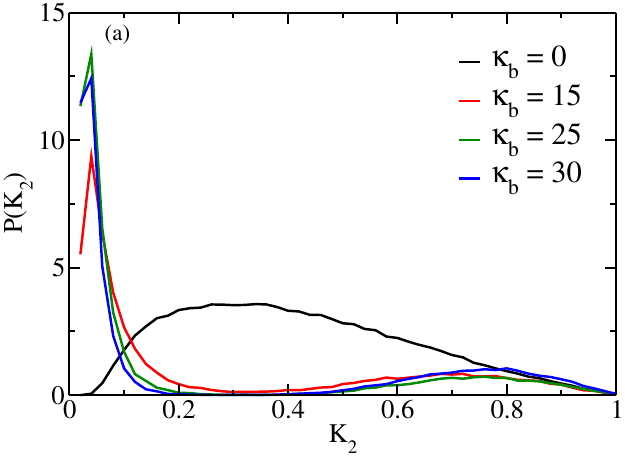}}
{\includegraphics[width=0.48\columnwidth]{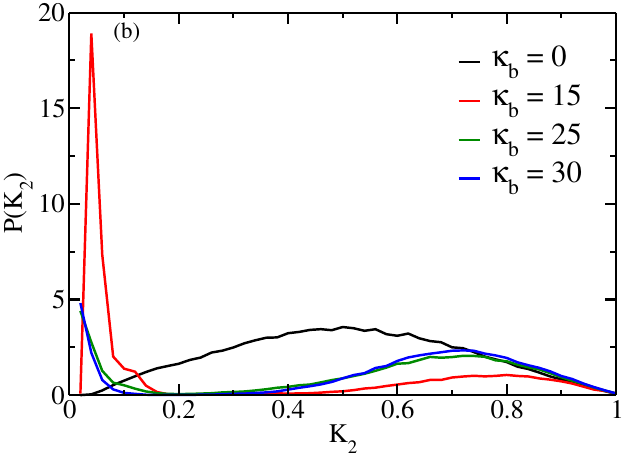}}
{\includegraphics[width=0.48\columnwidth]{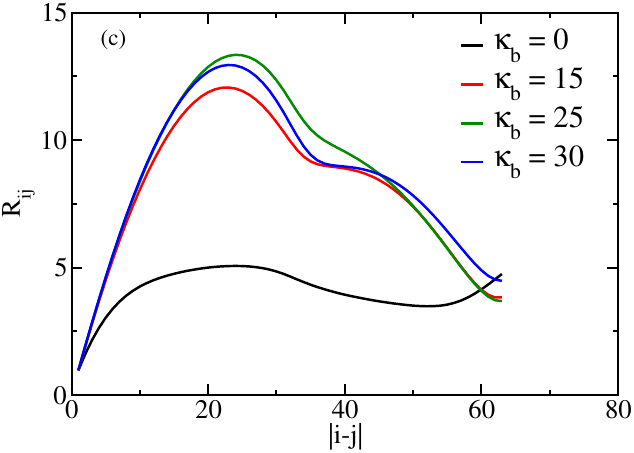}}
{\includegraphics[width=0.48\columnwidth]{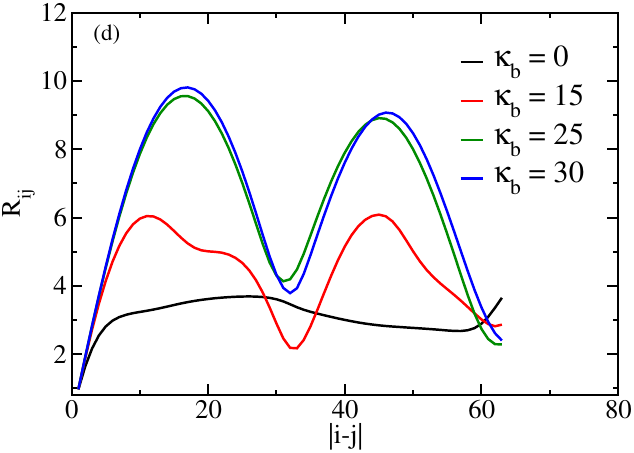}}
\caption{ The normalized probability distribution of the shape factor ($K_2$) and $R_{ij}$ for the case of diblock PA at different bending rigidities at $\Gamma_e=1$ and $\Gamma_e=6$. }
\label{Fig:space_dist_lb}
\end{figure}

\subsubsection{Shape of Diblock Chain}
We employ shape factors to differentiate various conformational phases of the PA, which serves as a valuable metric for characterizing deviations from spherically symmetric globular conformations towards planner, elongated, or extended states.  The shape factor\cite{khalatur1980effect} is defined as follows:

\begin{equation}
    K_1 = \left <\frac{\lambda_1}{\lambda_3}\right>  ; K_2 =  \left <\frac{\lambda_2}{\lambda_3}\right> ,
\end{equation}
where, $\lambda_1, \lambda_2,$ and $ \lambda_3$ correspond to the eigenvalues of the instantaneous gyration tensor.  Here, $\lambda_3$ and $\lambda_1$ are the  largest  and  smallest eigenvalues of  gyration tensor, respectively. 

For example, $K_2<1/2$ signifies an elongated conformation of the PA chain with smaller values of $K_1$.  $K_2 \approx 1/2, K_1\approx 1/2$ can be inferred as an elongated globular chain.  The $K_2>1/2$ range combined with a smaller value of $K_1<1/2$ indicates planner conformations, which may be circular or oval-shaped PA chain conformations.

The normalized probability distribution of shape factors ($K_2$) for diblock PA is depicted in the panels of Fig.~\ref{Fig:space_dist_lb} (a) and (b).  The probability distribution $P(K_2)$ reveals three distinct peaks nearly at $0.05$, $0.5$, and $0.7$ at various bending rigidities, indicating the presence of various stable conformational states.  The peak  $K_2\approx0.5$ is primarily observed for elliptical globular states; on the other hand, at $0.05$ corresponds to the bundle and hairpin-like state, and a peak near appearing at $0.7$ is the prevalence of a circular state for the diblock PA. 
{\cblue The most probable conformational state of the PA chain is identified from the cumulative sum of the normalized probability distribution of $K_2$ over specific ranges. The range between 0 to 0.15 corresponds to the hairpin and bundle shape conformations, while for the circular conformation, the relevant range lies between 0.4 to 1.0. Fig.\ref{Fig:space_dist_lb} (a) the cumulative sum within the 0 to 0.15 range is significantly higher for bending rigidity values of $\kappa_b = 15$, 25, and 30. This signifies that for $\Gamma_e = 1$, the hairpin configuration is the most probable state, but for Fig.~\ref{Fig:space_dist_lb} (b), the cumulative sum within the 0.4 to 1.0 range is notably larger for $\kappa_b = 25$ and 30, pointing to the circular conformation as the most probable state in this regime.  Notably, the height of the peak near $0.1$ is higher than the circular conformations, yet the cumulative sum is smaller than that at $\kappa_b=25$, and $30$. 
 }

Furthermore, we compute  $R_{ij}$, the ensemble-averaged distance between $i^{th}$ and $j^{th}$ monomers within the PA chain. The $R_{ij}$ profile offers insights into the spatial arrangements of monomers and their proximity to one another along the chain; see Figure~\ref{Fig:space_dist_lb} (c) and (d). 
The $R_{ij}$ grows with $|i-j|$ up to half chain length for all the bending rigidities due to electrostatic repulsion among alike charged residues. Beyond that, it weakly decreases at small bending parameters, i.e., $\kappa_b<4$ at $\Gamma_e=1$ as Fig.\ref{Fig:space_dist_lb}-c depicts. The decrease of $R_{ij}$ is more pronounced for larger $\kappa_b$ due to the bending of the semiflexible. In the intermediate bending regime, $ R_{ij}$ profiles exhibit two peaks of different heights, Fig.~\ref{Fig:space_dist_lb}-c. This indicates a more complex spatial arrangement of monomers within the chain. Further increments in $\kappa_b$ lead to the emergence of single peaks in the $R_{ij}$ profile, signaling the presence of a hairpin-like state or a folded structure from the center of the PA.  

In contrast, for $\Gamma_e=6$, the $R_{ij}$ profile exhibits two distinct peaks in the intermediate regime $10<\kappa_b<20$, indicating a higher probability of multiple folded states of the chain. However, as $\kappa_b$ rises, the $R_{ij}$ profile manifests two distinct peaks of approximately equal height in the range of $20\leq\kappa_b<30$, see Fig.\ref{Fig:space_dist_lb}-d. This profile suggests the presence of circular conformations. Notably, this observation is corroborated by the distribution plot of $K_2$, further accentuating the presence of the circular conformations within the PA chain.  For various values of $\Gamma_e$, the normalized distribution function $P(K_2)$ is displayed in SI-Fig.5.

\begin{figure}[tb]
\includegraphics[width=\columnwidth]{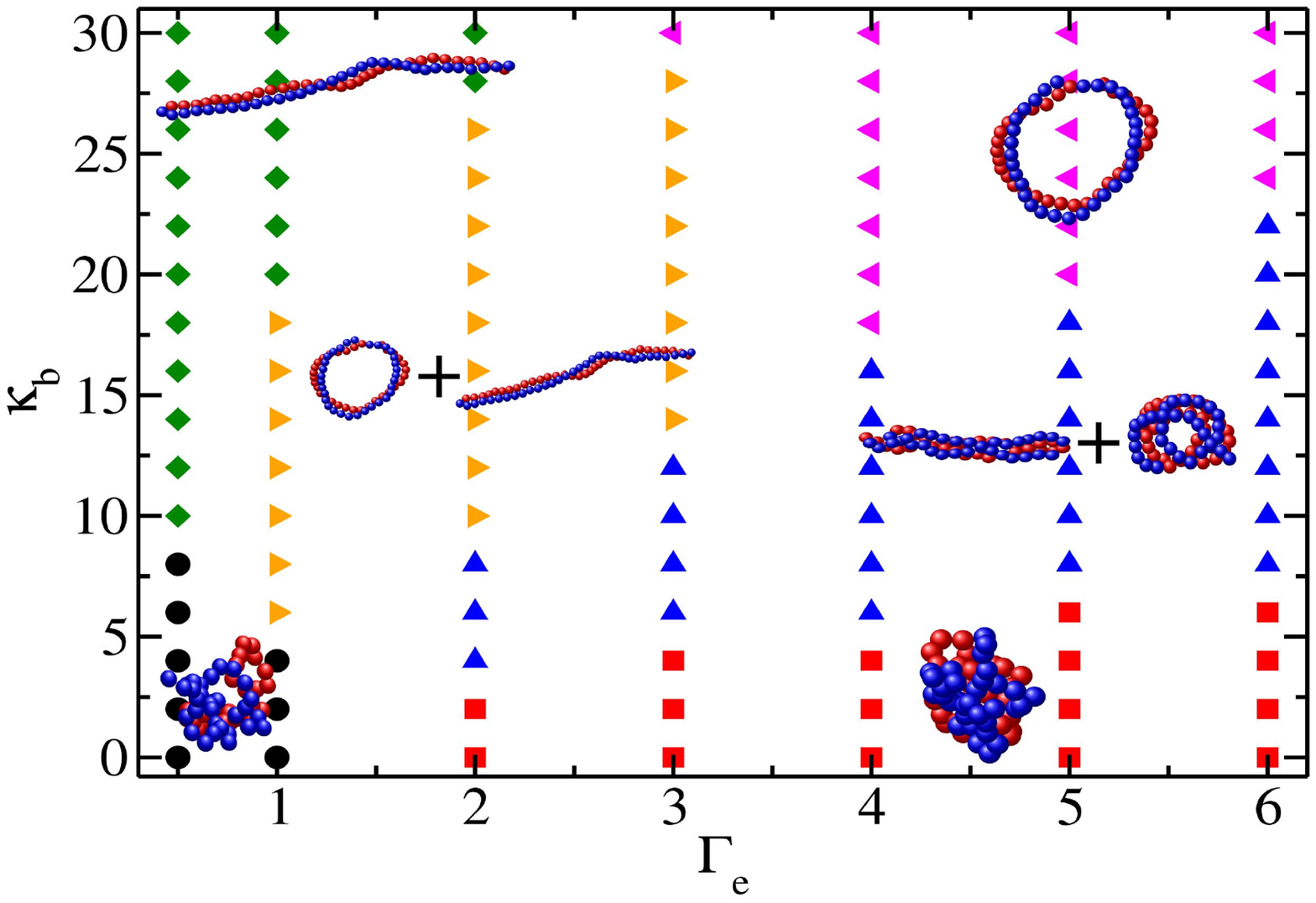}
\caption{{\cblue The phase diagram of the diblock PA chain's conformation in the parameter spaces $\Gamma_e$ and $\kappa_b$.  Various symbols correspond to different phases given, such as bullet (coil), square (globule), diamond (hairpin), triangle-left (circular),  triangle-up (mixed phase of folded and torus/elongated globule), and triangle-right (mixed phase of hairpin and circular states).}  }
\label{Fig:phase_vs_lb}
\end{figure}

\subsubsection{ Phase diagram}
Summarizing the results for the case of the diblock PA chain, we have constructed the phase diagram in Fig.~\ref{Fig:phase_vs_lb} in terms of bending rigidity $\kappa_b$ and electrostatic strength $\Gamma_e$. Figure~\ref{Fig:phase_vs_lb} visually illustrates the distinct states that emerge with higher probabilities across varying values of $\kappa_b$ and $\Gamma_e$. These transition points are determined from the probability distribution function $P(K_2)$ of the shape factor $K_2$. 

At lower $\Gamma_e$ values, an increase in $\kappa_b$ leads to a transition from a coil-like (bullets) state to hairpin-like (diamonds) conformations. However, for $\Gamma_e = 1,2,$ and $3$, the diblock PA chain adopts an elongated globule (filled square) conformations at lower $\kappa_b$ values. {\cblue Upon increasing  $\kappa_b$, a transition from a globular to a folded-state mixed with the torus and elongated globular states appears  (triangle up). This is  
 followed by coexistence of circular and hairpin states (triangle right), and finally hairpin-like (diamonds) or circular state appears (triangle left, for $\Gamma_e=3$) for larger $\kappa_b$.}    Conversely, at higher $\Gamma_e=4,5,$ and 6, a similar transition from globular to folded states is also observed, as for the case of lower $\Gamma_e$. However, with further increases in $\kappa_b$, the diblock PA chain predominantly assumes a circular conformation (triangle left).

\begin{figure}
{\includegraphics[width=\columnwidth]{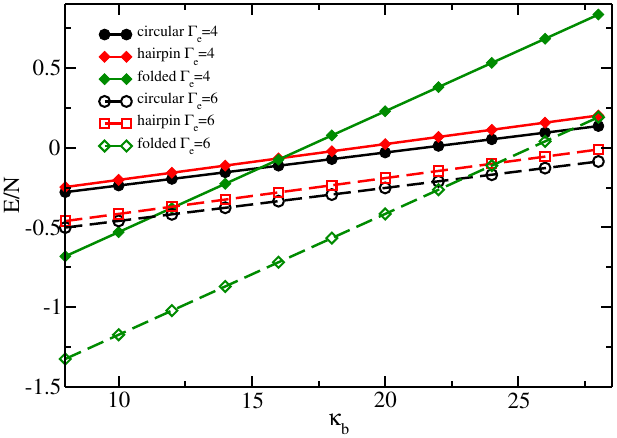}}
\caption{ Sum of electrostatic and bending energies for hair-pin (see Fig.\ref{Fig:diblock_conformation}-E), four-folded (see Fig.\ref{Fig:diblock_conformation}-C), and circular conformations (see Fig.\ref{Fig:diblock_conformation}-D) for various bending strengths for the case of diblock at $\Gamma_e=$ $4$ (filled symbols) and $6$ (open symbols).}
\label{Fig:energies_lb}
\end{figure}

 \subsubsection{ Energy } To investigate the stability of the various states and the transition between folded/hairpin and circular states, we also examine the energy variation across different conformational states. The bending  ($U_B$) and electrostatic energies ($U_C$) as a function of the bending parameter $\kappa_b$ for these conformations are presented in Fig.~\ref{Fig:energies_lb}. The total energy of the hairpin state (Fig.~\ref{Fig:diblock_conformation}-E), four-folded elongated conformations (Fig.~\ref{Fig:diblock_conformation}-C), and circular states (Fig.~\ref{Fig:diblock_conformation})-D) are plotted in Fig.~\ref{Fig:energies_lb} against $\kappa_b$ for the case of diblock at $\Gamma_e =$ $4$ and $6$. For $\Gamma_e=6$, at lower values of $\kappa_b$, the energy of four-folded elongated states is lower than that of the hairpin and circular states. However, nearly $\kappa_b\approx25$, there is a crossover in the energies of circular and hairpin states over the four-folded conformations. The plot indicates the transition from elongated to circular states also occurs at $\kappa_b\approx25$, as evident from Fig.~\ref{Fig:space_dist_lb}. 
More importantly, the energy of the hairpin state is higher than the circular states and lower than the elongated state; beyond $\kappa_b\approx25$, however, the difference in the energies of the circular states and the hairpin is very small; thereby, PA can acquire any of these states, see Fig.~\ref{Fig:energies_lb}. However, the circular states are more stable in this regime. 
For $\Gamma_e=4$, the transition point from the folded state to the circular state shifts to a lower value of $\kappa_b\approx 15$, which is also observed in the simulations, see Fig.\ref{Fig:phase_vs_lb}.

The circular states and their curvature radius can also be quantified by the tangent-tangent correlations  ($C(s)$) and radial distribution function  ($g(r)$) of the chain. The correlations and radial distribution functions are shown in the SI-Figures 2 and 3. Further, we also show that the circular states also appear for the longer chain lengths ($N=256$); see SI-Fig-2a and SI-Fig-2b.

The transition between various conformational states is driven by the electrostatic attraction and bending penalties. In the regime of weak bending penalty ($\kappa_b$), the electrostatic attractions over the bending dominate; therefore, the chain acquires folded and globular conformations. However, in the intermediate range, a considerable bending penalty favors the chain's adoption of a smoother bend along its contour, further complemented by electrostatic attractions, leading to stable circular states. Sharply bent conformations exhibit a larger bending penalty than the smoothly bent circular states; therefore, they appear as meta-stable states in the simulations. A rod-like or hair-pin state will appear in the regime of large bending.

\begin{figure}
{\includegraphics[width=0.49\columnwidth]{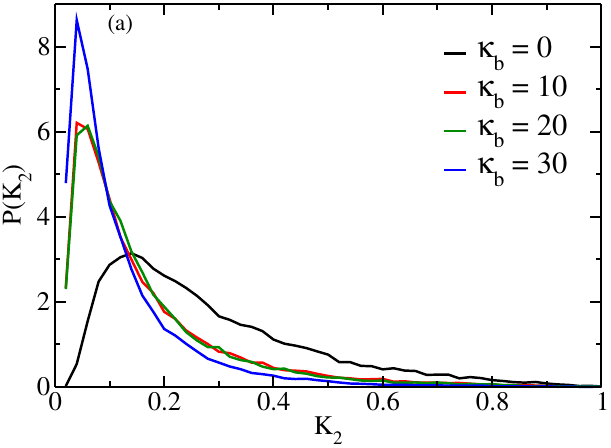}}
{\includegraphics[width=0.49\columnwidth]{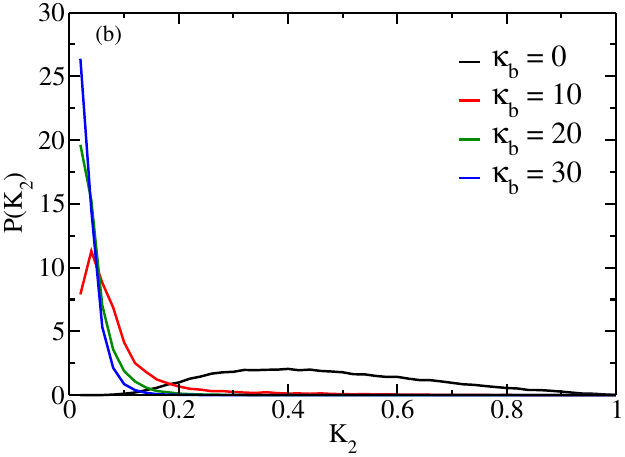}}
\caption{ The normalized probability distribution of the shape factor ($K_2$) for alternate (a, $L_B=1$), and hexadeca-block (b, $L_B=4$) at various bending rigidities at a given $\Gamma_e=5$ and chain length $N=64$. }
\label{Fig:space_dist}
\end{figure}

\subsection{Conformations of Smaller Blocks Lengths}
The previous section discussed the structural transitions of a semiflexible diblock (largest block length) PA chain.  In this section, we consider the conformations of the smaller block lengths,  alternating charge sequence  $L_B=1$ and $L_B=4$. 
 The probability distribution of shape factor ($K_2$) for varying block length is depicted in the panels of Fig.~\ref{Fig:space_dist} (a) and (b), encompassing hexadeca-block ($L_B=4$), and alternating ($L_B=1$) sequences for different bending rigidities. \\
 
 {\cblue Figure~\ref{Fig:space_dist}-a displays that the $P(K_2)$ of alternating charge sequence has one peak, which shifts towards the left and slowly decays to zero. For larger bending rigidities, the PA chain acquires rod-like conformations. While, for block length 4, the peak for $\kappa_b = 0$ is at 0.4, which suggests the existence of globular conformation. With the increase in $\kappa_b$, the peak shifts to the left, and it decays quickly to zero,  indicating the existence of hairpin-like conformations for larger bending rigidities as observed for the diblock PA for lower $\Gamma_e$, see Fig.\ref{Fig:phase_vs_lb}. 
}

\begin{figure}
{\includegraphics[width=0.49\columnwidth]{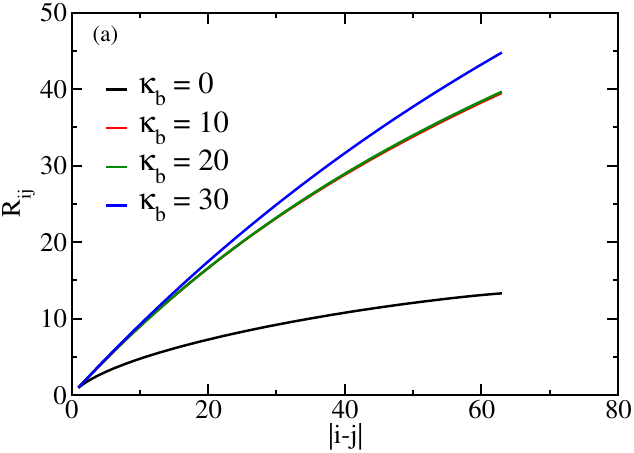}}
{\includegraphics[width=0.49\columnwidth]{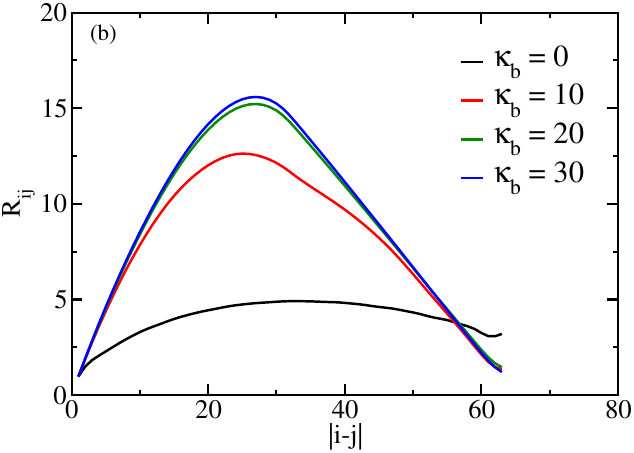}}
\caption{ The $R_{ij}$ profile for alternate   $L_B=1$(a), and hexadeca-block $L_B=4$(b) sequences  at various bending rigidities for $\Gamma_e=5$ and  $N=64$. }
\label{Fig:rij_lb5}
\end{figure}

Further, we examine the distance between various monomers ($R_{ij}$) in the chain, as illustrated in Fig.~\ref{Fig:rij_lb5}. For an alternating charged sequence, as expected,  the $R_{ij}$ profiles Fig.~\ref{Fig:rij_lb5}-a displays a monotonic increasing trend for all $\kappa_b$.  $R_{ij}$ increases rapidly for higher bending, suggesting a continuous structural transition from a coil-like to an elongated semi-flexible PA chain. 
{\cblue For block length $4$, the $R_{ij}$ profile increases nearly up to half chain length $|i-j| \approx32 $; beyond that, it weakly decreases further for large distances. The height of the peak further increases with an increase in $\kappa_b$, which suggests that the PA chain has more stretched conformation. For larger $\kappa_b$ and beyond $|i-j| \approx32 $, the profile decreases and exhibits a symmetry around the maximum.   The emergence of a single peak and symmetry in the $R_{ij}$ profile indicates the formation of a hairpin-like conformation which is consistent with the observations from the probability distribution of the $K_2$ as shown Fig.~\ref{Fig:space_dist}.}



\section{Conclusion}
In this article, we have presented a systematic study of the structural transition of flexible and semi-flexible PA chains in salt-free solutions with the help of coarse-grained molecular dynamics simulations.  The block length of the charged residues from segregated to homogeneously mixed sequence, and the bending rigidity of the PA are varied.  Our study reveals that the flexible PA undergoes a coil-globule transition depending on the variation of the charge sequence (nSCD).  The homogeneously mixed charge-sequence PA exhibits the coil state, which undergoes a continuous transition to a globular state for the charge-segregated state.  
Additionally, we have shown that the universal behavior of the radius of gyration of the PA chain is achieved once it's scaled with the $\theta$-point values for various electrostatic interaction strengths. The stronger electrostatic interaction among larger block lengths drives the transition from the coil to the globular conformational stats. 

Moreover, we have shown the dynamical behavior of the PA chain by computing the MSD of a monomer, which displays that the internal dynamics of the various monomers significantly differ in larger block PA to a homogeneously mixed states. The obtained subdiffusive behavior in the intermediate time regime has an exponent: $\beta\approx1/2$ for the $\theta$-point and the globular states; the exponent $\beta\approx3/5$ for the homogeneously mixed state without HI. In addition, our study reveals that HI has a crucial role in the internal dynamics of the PA; this leads to the sub-diffusive exponents being modified in the presence of the solvent-mediated interaction from $\beta\approx 3/5$ to $2/3$  for the case of the homogeneously mixed state and $\beta\approx1/2$ to $3/5$ for the globular states and  $\theta$-point sequence.\cite{doi:86,Muthu_Book_2011}

 In addition to exploring flexible polyampholyte (PA) chains,  we have also emphasized the structural behavior of a semi-flexible PA chain,  which displays intriguing structural transitions based on the variation in bending rigidity. These transitions result in diverse conformations, such as globular, coil-state, bundle-shape, circular/torus, and rod-like structures, contingent upon the sequence of charged residues and electrostatic interaction. The shift between these states is influenced by the interplay between long-range electrostatic interactions among the charged residues and the bending stiffness of the chain, which seeks to elongate the PA chain. Our findings reveal that the circular and hairpin states for the intermediate bending regime are stable conformations for PA chains, especially those with larger nSCD values (larger block lengths). Conversely, PA chains with lower nSCD values exhibit stable coil- and rod-like conformations. 
The radial distribution function and tangent-vector correlation are adequate tools for locating the circular state of the chain, providing insights into its average radius, see SI-Fig-2 and 3. Small shape factors and relatively high electrostatic and bending energies characterize the folded and rod-like states. Such structural transitions have been revealed for the neutral and dipolar chains based on the variation of temperature and electrostatic interaction strengths.\cite{aierken2023stable,pantawane2022temperature,gordievskaya2019conformational}

In conclusion, our study offers valuable insights into the conformational dynamics of polyampholytic chains characterized by periodic charge sequences, electrostatic interaction parameters, and bending rigidities. Our work enhances understanding of the conformational patterns exhibited by semi-flexible intrinsically disordered proteins (IDPs) and biopolymers. It would be more plausible to consider the specific proteins and probe the structural and transport behavior into various parameter regimes.

\section{Supplementary Material}
The supplementary text provides a brief overview of the MPC method and its simulation parameters.  In addition, it includes the tangent correlation function, radial distribution function, shape factors, and their distribution function, along with a description of supporting movie files.

\section{Acknowledgements}
SPS and RP acknowledge financial support from the DST-SERB Grant No. CRG/2020/000661 and UGC, respectively. The computational facilities at IISER Bhopal are highly acknowledged for providing computational time.



\begin{thebibliography}{75}%
\makeatletter
\providecommand \@ifxundefined [1]{%
 \@ifx{#1\undefined}
}%
\providecommand \@ifnum [1]{%
 \ifnum #1\expandafter \@firstoftwo
 \else \expandafter \@secondoftwo
 \fi
}%
\providecommand \@ifx [1]{%
 \ifx #1\expandafter \@firstoftwo
 \else \expandafter \@secondoftwo
 \fi
}%
\providecommand \natexlab [1]{#1}%
\providecommand \enquote  [1]{``#1''}%
\providecommand \bibnamefont  [1]{#1}%
\providecommand \bibfnamefont [1]{#1}%
\providecommand \citenamefont [1]{#1}%
\providecommand \href@noop [0]{\@secondoftwo}%
\providecommand \href [0]{\begingroup \@sanitize@url \@href}%
\providecommand \@href[1]{\@@startlink{#1}\@@href}%
\providecommand \@@href[1]{\endgroup#1\@@endlink}%
\providecommand \@sanitize@url [0]{\catcode `\\12\catcode `\$12\catcode
  `\&12\catcode `\#12\catcode `\^12\catcode `\_12\catcode `\%12\relax}%
\providecommand \@@startlink[1]{}%
\providecommand \@@endlink[0]{}%
\providecommand \url  [0]{\begingroup\@sanitize@url \@url }%
\providecommand \@url [1]{\endgroup\@href {#1}{\urlprefix }}%
\providecommand \urlprefix  [0]{URL }%
\providecommand \Eprint [0]{\href }%
\providecommand \doibase [0]{https://doi.org/}%
\providecommand \selectlanguage [0]{\@gobble}%
\providecommand \bibinfo  [0]{\@secondoftwo}%
\providecommand \bibfield  [0]{\@secondoftwo}%
\providecommand \translation [1]{[#1]}%
\providecommand \BibitemOpen [0]{}%
\providecommand \bibitemStop [0]{}%
\providecommand \bibitemNoStop [0]{.\EOS\space}%
\providecommand \EOS [0]{\spacefactor3000\relax}%
\providecommand \BibitemShut  [1]{\csname bibitem#1\endcsname}%
\let\auto@bib@innerbib\@empty
\bibitem [{\citenamefont {Lowe}\ and\ \citenamefont
  {McCormick}(2002)}]{lowe2002synthesis}%
  \BibitemOpen
  \bibfield  {author} {\bibinfo {author} {\bibfnamefont {A.~B.}\ \bibnamefont
  {Lowe}}\ and\ \bibinfo {author} {\bibfnamefont {C.~L.}\ \bibnamefont
  {McCormick}},\ }\bibfield  {title} {\enquote {\bibinfo {title} {Synthesis and
  solution properties of zwitterionic polymers},}\ }\href@noop {} {\bibfield
  {journal} {\bibinfo  {journal} {Chemical reviews}\ }\textbf {\bibinfo
  {volume} {102}},\ \bibinfo {pages} {4177--4190} (\bibinfo {year}
  {2002})}\BibitemShut {NoStop}%
\bibitem [{\citenamefont {Dobrynin}\ \emph {et~al.}(2004)\citenamefont
  {Dobrynin}, \citenamefont {Colby}, \citenamefont {Rubinstein},\ and\
  \citenamefont {Polyampholates}}]{dobrynin2004polym}%
  \BibitemOpen
  \bibfield  {author} {\bibinfo {author} {\bibfnamefont {A.}~\bibnamefont
  {Dobrynin}}, \bibinfo {author} {\bibfnamefont {R.}~\bibnamefont {Colby}},
  \bibinfo {author} {\bibfnamefont {M.}~\bibnamefont {Rubinstein}},\ and\
  \bibinfo {author} {\bibfnamefont {J.}~\bibnamefont {Polyampholates}},\
  }\bibfield  {title} {\enquote {\bibinfo {title} {of polym},}\ }\href@noop {}
  {\bibfield  {journal} {\bibinfo  {journal} {Sci.: Part B: Polymer Physics}\
  }\textbf {\bibinfo {volume} {42}},\ \bibinfo {pages} {3513--3538} (\bibinfo
  {year} {2004})}\BibitemShut {NoStop}%
\bibitem [{\citenamefont {Higgs}\ and\ \citenamefont
  {Joanny}(1991)}]{higgs1991theory}%
  \BibitemOpen
  \bibfield  {author} {\bibinfo {author} {\bibfnamefont {P.~G.}\ \bibnamefont
  {Higgs}}\ and\ \bibinfo {author} {\bibfnamefont {J.-F.}\ \bibnamefont
  {Joanny}},\ }\bibfield  {title} {\enquote {\bibinfo {title} {Theory of
  polyampholyte solutions},}\ }\href@noop {} {\bibfield  {journal} {\bibinfo
  {journal} {The Journal of chemical physics}\ }\textbf {\bibinfo {volume}
  {94}},\ \bibinfo {pages} {1543--1554} (\bibinfo {year} {1991})}\BibitemShut
  {NoStop}%
\bibitem [{\citenamefont {Silmore}\ and\ \citenamefont
  {Kumar}(2021)}]{silmore2021dynamics}%
  \BibitemOpen
  \bibfield  {author} {\bibinfo {author} {\bibfnamefont {K.~S.}\ \bibnamefont
  {Silmore}}\ and\ \bibinfo {author} {\bibfnamefont {R.}~\bibnamefont
  {Kumar}},\ }\bibfield  {title} {\enquote {\bibinfo {title} {Dynamics of a
  single polyampholyte chain},}\ }\href@noop {} {\bibfield  {journal} {\bibinfo
   {journal} {The Journal of Chemical Physics}\ }\textbf {\bibinfo {volume}
  {155}},\ \bibinfo {pages} {214903} (\bibinfo {year} {2021})}\BibitemShut
  {NoStop}%
\bibitem [{\citenamefont {Dignon}, \citenamefont {Best},\ and\ \citenamefont
  {Mittal}(2020)}]{dignon2020biomolecular}%
  \BibitemOpen
  \bibfield  {author} {\bibinfo {author} {\bibfnamefont {G.~L.}\ \bibnamefont
  {Dignon}}, \bibinfo {author} {\bibfnamefont {R.~B.}\ \bibnamefont {Best}},\
  and\ \bibinfo {author} {\bibfnamefont {J.}~\bibnamefont {Mittal}},\
  }\bibfield  {title} {\enquote {\bibinfo {title} {Biomolecular phase
  separation: from molecular driving forces to macroscopic properties},}\
  }\href@noop {} {\bibfield  {journal} {\bibinfo  {journal} {Annual review of
  physical chemistry}\ }\textbf {\bibinfo {volume} {71}},\ \bibinfo {pages}
  {53--75} (\bibinfo {year} {2020})}\BibitemShut {NoStop}%
\bibitem [{\citenamefont {Das}\ and\ \citenamefont
  {Pappu}(2013)}]{das2013conformations}%
  \BibitemOpen
  \bibfield  {author} {\bibinfo {author} {\bibfnamefont {R.~K.}\ \bibnamefont
  {Das}}\ and\ \bibinfo {author} {\bibfnamefont {R.~V.}\ \bibnamefont
  {Pappu}},\ }\bibfield  {title} {\enquote {\bibinfo {title} {Conformations of
  intrinsically disordered proteins are influenced by linear sequence
  distributions of oppositely charged residues},}\ }\href@noop {} {\bibfield
  {journal} {\bibinfo  {journal} {Proceedings of the National Academy of
  Sciences}\ }\textbf {\bibinfo {volume} {110}},\ \bibinfo {pages}
  {13392--13397} (\bibinfo {year} {2013})}\BibitemShut {NoStop}%
\bibitem [{\citenamefont {Long}\ \emph {et~al.}(1998)\citenamefont {Long},
  \citenamefont {Dobrynin}, \citenamefont {Rubinstein},\ and\ \citenamefont
  {Ajdari}}]{long1998electrophoresis}%
  \BibitemOpen
  \bibfield  {author} {\bibinfo {author} {\bibfnamefont {D.}~\bibnamefont
  {Long}}, \bibinfo {author} {\bibfnamefont {A.~V.}\ \bibnamefont {Dobrynin}},
  \bibinfo {author} {\bibfnamefont {M.}~\bibnamefont {Rubinstein}},\ and\
  \bibinfo {author} {\bibfnamefont {A.}~\bibnamefont {Ajdari}},\ }\bibfield
  {title} {\enquote {\bibinfo {title} {Electrophoresis of polyampholytes},}\
  }\href@noop {} {\bibfield  {journal} {\bibinfo  {journal} {The Journal of
  chemical physics}\ }\textbf {\bibinfo {volume} {108}},\ \bibinfo {pages}
  {1234--1244} (\bibinfo {year} {1998})}\BibitemShut {NoStop}%
\bibitem [{\citenamefont {Rumyantsev}, \citenamefont {Johner},\ and\
  \citenamefont {de~Pablo}(2021)}]{rumyantsev2021sequence}%
  \BibitemOpen
  \bibfield  {author} {\bibinfo {author} {\bibfnamefont {A.~M.}\ \bibnamefont
  {Rumyantsev}}, \bibinfo {author} {\bibfnamefont {A.}~\bibnamefont {Johner}},\
  and\ \bibinfo {author} {\bibfnamefont {J.~J.}\ \bibnamefont {de~Pablo}},\
  }\bibfield  {title} {\enquote {\bibinfo {title} {Sequence blockiness controls
  the structure of polyampholyte necklaces},}\ }\href@noop {} {\bibfield
  {journal} {\bibinfo  {journal} {ACS Macro Letters}\ }\textbf {\bibinfo
  {volume} {10}},\ \bibinfo {pages} {1048--1054} (\bibinfo {year}
  {2021})}\BibitemShut {NoStop}%
\bibitem [{\citenamefont {Dyson}\ and\ \citenamefont
  {Wright}(2005)}]{dyson2005intrinsically}%
  \BibitemOpen
  \bibfield  {author} {\bibinfo {author} {\bibfnamefont {H.~J.}\ \bibnamefont
  {Dyson}}\ and\ \bibinfo {author} {\bibfnamefont {P.~E.}\ \bibnamefont
  {Wright}},\ }\bibfield  {title} {\enquote {\bibinfo {title} {Intrinsically
  unstructured proteins and their functions},}\ }\href@noop {} {\bibfield
  {journal} {\bibinfo  {journal} {Nature reviews Molecular cell biology}\
  }\textbf {\bibinfo {volume} {6}},\ \bibinfo {pages} {197--208} (\bibinfo
  {year} {2005})}\BibitemShut {NoStop}%
\bibitem [{\citenamefont {Tantos}, \citenamefont {Han},\ and\ \citenamefont
  {Tompa}(2012)}]{tantos2012intrinsic}%
  \BibitemOpen
  \bibfield  {author} {\bibinfo {author} {\bibfnamefont {A.}~\bibnamefont
  {Tantos}}, \bibinfo {author} {\bibfnamefont {K.-H.}\ \bibnamefont {Han}},\
  and\ \bibinfo {author} {\bibfnamefont {P.}~\bibnamefont {Tompa}},\ }\bibfield
   {title} {\enquote {\bibinfo {title} {Intrinsic disorder in cell signaling
  and gene transcription},}\ }\href@noop {} {\bibfield  {journal} {\bibinfo
  {journal} {Molecular and cellular endocrinology}\ }\textbf {\bibinfo {volume}
  {348}},\ \bibinfo {pages} {457--465} (\bibinfo {year} {2012})}\BibitemShut
  {NoStop}%
\bibitem [{\citenamefont {Wright}\ and\ \citenamefont
  {Dyson}(2015)}]{wright2015intrinsically}%
  \BibitemOpen
  \bibfield  {author} {\bibinfo {author} {\bibfnamefont {P.~E.}\ \bibnamefont
  {Wright}}\ and\ \bibinfo {author} {\bibfnamefont {H.~J.}\ \bibnamefont
  {Dyson}},\ }\bibfield  {title} {\enquote {\bibinfo {title} {Intrinsically
  disordered proteins in cellular signalling and regulation},}\ }\href@noop {}
  {\bibfield  {journal} {\bibinfo  {journal} {Nature reviews Molecular cell
  biology}\ }\textbf {\bibinfo {volume} {16}},\ \bibinfo {pages} {18--29}
  (\bibinfo {year} {2015})}\BibitemShut {NoStop}%
\bibitem [{\citenamefont {Dima}\ and\ \citenamefont
  {Thirumalai}(2004)}]{dima2004proteins}%
  \BibitemOpen
  \bibfield  {author} {\bibinfo {author} {\bibfnamefont {R.~I.}\ \bibnamefont
  {Dima}}\ and\ \bibinfo {author} {\bibfnamefont {D.}~\bibnamefont
  {Thirumalai}},\ }\bibfield  {title} {\enquote {\bibinfo {title} {Proteins
  associated with diseases show enhanced sequence correlation between charged
  residues},}\ }\href@noop {} {\bibfield  {journal} {\bibinfo  {journal}
  {Bioinformatics}\ }\textbf {\bibinfo {volume} {20}},\ \bibinfo {pages}
  {2345--2354} (\bibinfo {year} {2004})}\BibitemShut {NoStop}%
\bibitem [{\citenamefont {Shin}\ and\ \citenamefont
  {Brangwynne}(2017)}]{shin2017liquid}%
  \BibitemOpen
  \bibfield  {author} {\bibinfo {author} {\bibfnamefont {Y.}~\bibnamefont
  {Shin}}\ and\ \bibinfo {author} {\bibfnamefont {C.~P.}\ \bibnamefont
  {Brangwynne}},\ }\bibfield  {title} {\enquote {\bibinfo {title} {Liquid phase
  condensation in cell physiology and disease},}\ }\href@noop {} {\bibfield
  {journal} {\bibinfo  {journal} {Science}\ }\textbf {\bibinfo {volume}
  {357}},\ \bibinfo {pages} {eaaf4382} (\bibinfo {year} {2017})}\BibitemShut
  {NoStop}%
\bibitem [{\citenamefont {Banani}\ \emph {et~al.}(2017)\citenamefont {Banani},
  \citenamefont {Lee}, \citenamefont {Hyman},\ and\ \citenamefont
  {Rosen}}]{banani2017biomolecular}%
  \BibitemOpen
  \bibfield  {author} {\bibinfo {author} {\bibfnamefont {S.~F.}\ \bibnamefont
  {Banani}}, \bibinfo {author} {\bibfnamefont {H.~O.}\ \bibnamefont {Lee}},
  \bibinfo {author} {\bibfnamefont {A.~A.}\ \bibnamefont {Hyman}},\ and\
  \bibinfo {author} {\bibfnamefont {M.~K.}\ \bibnamefont {Rosen}},\ }\bibfield
  {title} {\enquote {\bibinfo {title} {Biomolecular condensates: organizers of
  cellular biochemistry},}\ }\href@noop {} {\bibfield  {journal} {\bibinfo
  {journal} {Nature reviews Molecular cell biology}\ }\textbf {\bibinfo
  {volume} {18}},\ \bibinfo {pages} {285--298} (\bibinfo {year}
  {2017})}\BibitemShut {NoStop}%
\bibitem [{\citenamefont {Brangwynne}\ \emph {et~al.}(2009)\citenamefont
  {Brangwynne}, \citenamefont {Eckmann}, \citenamefont {Courson}, \citenamefont
  {Rybarska}, \citenamefont {Hoege}, \citenamefont {Gharakhani}, \citenamefont
  {J{\"u}licher},\ and\ \citenamefont {Hyman}}]{brangwynne2009germline}%
  \BibitemOpen
  \bibfield  {author} {\bibinfo {author} {\bibfnamefont {C.~P.}\ \bibnamefont
  {Brangwynne}}, \bibinfo {author} {\bibfnamefont {C.~R.}\ \bibnamefont
  {Eckmann}}, \bibinfo {author} {\bibfnamefont {D.~S.}\ \bibnamefont
  {Courson}}, \bibinfo {author} {\bibfnamefont {A.}~\bibnamefont {Rybarska}},
  \bibinfo {author} {\bibfnamefont {C.}~\bibnamefont {Hoege}}, \bibinfo
  {author} {\bibfnamefont {J.}~\bibnamefont {Gharakhani}}, \bibinfo {author}
  {\bibfnamefont {F.}~\bibnamefont {J{\"u}licher}},\ and\ \bibinfo {author}
  {\bibfnamefont {A.~A.}\ \bibnamefont {Hyman}},\ }\bibfield  {title} {\enquote
  {\bibinfo {title} {Germline p granules are liquid droplets that localize by
  controlled dissolution/condensation},}\ }\href@noop {} {\bibfield  {journal}
  {\bibinfo  {journal} {Science}\ }\textbf {\bibinfo {volume} {324}},\ \bibinfo
  {pages} {1729--1732} (\bibinfo {year} {2009})}\BibitemShut {NoStop}%
\bibitem [{\citenamefont {Bright}, \citenamefont {Woolf},\ and\ \citenamefont
  {Hoh}(2001)}]{bright2001predicting}%
  \BibitemOpen
  \bibfield  {author} {\bibinfo {author} {\bibfnamefont {J.~N.}\ \bibnamefont
  {Bright}}, \bibinfo {author} {\bibfnamefont {T.~B.}\ \bibnamefont {Woolf}},\
  and\ \bibinfo {author} {\bibfnamefont {J.~H.}\ \bibnamefont {Hoh}},\
  }\bibfield  {title} {\enquote {\bibinfo {title} {Predicting properties of
  intrinsically unstructured proteins},}\ }\href@noop {} {\bibfield  {journal}
  {\bibinfo  {journal} {Progress in biophysics and molecular biology}\ }\textbf
  {\bibinfo {volume} {76}},\ \bibinfo {pages} {131--173} (\bibinfo {year}
  {2001})}\BibitemShut {NoStop}%
\bibitem [{\citenamefont {Jiang}\ \emph {et~al.}(2006)\citenamefont {Jiang},
  \citenamefont {Feng}, \citenamefont {Liu},\ and\ \citenamefont
  {Hu}}]{jiang2006phase}%
  \BibitemOpen
  \bibfield  {author} {\bibinfo {author} {\bibfnamefont {J.}~\bibnamefont
  {Jiang}}, \bibinfo {author} {\bibfnamefont {J.}~\bibnamefont {Feng}},
  \bibinfo {author} {\bibfnamefont {H.}~\bibnamefont {Liu}},\ and\ \bibinfo
  {author} {\bibfnamefont {Y.}~\bibnamefont {Hu}},\ }\bibfield  {title}
  {\enquote {\bibinfo {title} {Phase behavior of polyampholytes from charged
  hard-sphere chain model},}\ }\href@noop {} {\bibfield  {journal} {\bibinfo
  {journal} {The Journal of chemical physics}\ }\textbf {\bibinfo {volume}
  {124}} (\bibinfo {year} {2006})}\BibitemShut {NoStop}%
\bibitem [{\citenamefont {Castelnovo}\ and\ \citenamefont
  {Joanny}(2002)}]{castelnovo2002phase}%
  \BibitemOpen
  \bibfield  {author} {\bibinfo {author} {\bibfnamefont {M.}~\bibnamefont
  {Castelnovo}}\ and\ \bibinfo {author} {\bibfnamefont {J.}~\bibnamefont
  {Joanny}},\ }\bibfield  {title} {\enquote {\bibinfo {title} {Phase diagram of
  diblock polyampholyte solutions},}\ }\href@noop {} {\bibfield  {journal}
  {\bibinfo  {journal} {Macromolecules}\ }\textbf {\bibinfo {volume} {35}},\
  \bibinfo {pages} {4531--4538} (\bibinfo {year} {2002})}\BibitemShut {NoStop}%
\bibitem [{\citenamefont {Shusharina}\ \emph {et~al.}(2005)\citenamefont
  {Shusharina}, \citenamefont {Zhulina}, \citenamefont {Dobrynin},\ and\
  \citenamefont {Rubinstein}}]{shusharina2005scaling}%
  \BibitemOpen
  \bibfield  {author} {\bibinfo {author} {\bibfnamefont {N.}~\bibnamefont
  {Shusharina}}, \bibinfo {author} {\bibfnamefont {E.}~\bibnamefont {Zhulina}},
  \bibinfo {author} {\bibfnamefont {A.}~\bibnamefont {Dobrynin}},\ and\
  \bibinfo {author} {\bibfnamefont {M.}~\bibnamefont {Rubinstein}},\ }\bibfield
   {title} {\enquote {\bibinfo {title} {Scaling theory of diblock polyampholyte
  solutions},}\ }\href@noop {} {\bibfield  {journal} {\bibinfo  {journal}
  {Macromolecules}\ }\textbf {\bibinfo {volume} {38}},\ \bibinfo {pages}
  {8870--8881} (\bibinfo {year} {2005})}\BibitemShut {NoStop}%
\bibitem [{\citenamefont {Wang}\ and\ \citenamefont
  {Rubinstein}(2006)}]{wang2006regimes}%
  \BibitemOpen
  \bibfield  {author} {\bibinfo {author} {\bibfnamefont {Z.}~\bibnamefont
  {Wang}}\ and\ \bibinfo {author} {\bibfnamefont {M.}~\bibnamefont
  {Rubinstein}},\ }\bibfield  {title} {\enquote {\bibinfo {title} {Regimes of
  conformational transitions of a diblock polyampholyte},}\ }\href@noop {}
  {\bibfield  {journal} {\bibinfo  {journal} {Macromolecules}\ }\textbf
  {\bibinfo {volume} {39}},\ \bibinfo {pages} {5897--5912} (\bibinfo {year}
  {2006})}\BibitemShut {NoStop}%
\bibitem [{\citenamefont {Cheong}\ and\ \citenamefont
  {Panagiotopoulos*}(2005)}]{cheong2005phase}%
  \BibitemOpen
  \bibfield  {author} {\bibinfo {author} {\bibfnamefont {D.~W.}\ \bibnamefont
  {Cheong}}\ and\ \bibinfo {author} {\bibfnamefont {A.~Z.}\ \bibnamefont
  {Panagiotopoulos*}},\ }\bibfield  {title} {\enquote {\bibinfo {title} {Phase
  behaviour of polyampholyte chains from grand canonical monte carlo
  simulations},}\ }\href@noop {} {\bibfield  {journal} {\bibinfo  {journal}
  {Molecular Physics}\ }\textbf {\bibinfo {volume} {103}},\ \bibinfo {pages}
  {3031--3044} (\bibinfo {year} {2005})}\BibitemShut {NoStop}%
\bibitem [{\citenamefont {Everaers}, \citenamefont {Johner},\ and\
  \citenamefont {Joanny}(1997)}]{everaers1997complexation}%
  \BibitemOpen
  \bibfield  {author} {\bibinfo {author} {\bibfnamefont {R.}~\bibnamefont
  {Everaers}}, \bibinfo {author} {\bibfnamefont {A.}~\bibnamefont {Johner}},\
  and\ \bibinfo {author} {\bibfnamefont {J.-F.}\ \bibnamefont {Joanny}},\
  }\bibfield  {title} {\enquote {\bibinfo {title} {Complexation and
  precipitation in polyampholyte solutions},}\ }\href@noop {} {\bibfield
  {journal} {\bibinfo  {journal} {Europhysics Letters}\ }\textbf {\bibinfo
  {volume} {37}},\ \bibinfo {pages} {275} (\bibinfo {year} {1997})}\BibitemShut
  {NoStop}%
\bibitem [{\citenamefont {Barbosa}\ and\ \citenamefont
  {Levin}(1996)}]{barbosa1996phase}%
  \BibitemOpen
  \bibfield  {author} {\bibinfo {author} {\bibfnamefont {M.~C.}\ \bibnamefont
  {Barbosa}}\ and\ \bibinfo {author} {\bibfnamefont {Y.}~\bibnamefont
  {Levin}},\ }\bibfield  {title} {\enquote {\bibinfo {title} {Phase transitions
  of a neutral polyampholyte},}\ }\href@noop {} {\bibfield  {journal} {\bibinfo
   {journal} {Physica A: Statistical Mechanics and its Applications}\ }\textbf
  {\bibinfo {volume} {231}},\ \bibinfo {pages} {467--483} (\bibinfo {year}
  {1996})}\BibitemShut {NoStop}%
\bibitem [{\citenamefont {Mao}\ \emph {et~al.}(2010)\citenamefont {Mao},
  \citenamefont {Crick}, \citenamefont {Vitalis}, \citenamefont {Chicoine},\
  and\ \citenamefont {Pappu}}]{mao2010net}%
  \BibitemOpen
  \bibfield  {author} {\bibinfo {author} {\bibfnamefont {A.~H.}\ \bibnamefont
  {Mao}}, \bibinfo {author} {\bibfnamefont {S.~L.}\ \bibnamefont {Crick}},
  \bibinfo {author} {\bibfnamefont {A.}~\bibnamefont {Vitalis}}, \bibinfo
  {author} {\bibfnamefont {C.~L.}\ \bibnamefont {Chicoine}},\ and\ \bibinfo
  {author} {\bibfnamefont {R.~V.}\ \bibnamefont {Pappu}},\ }\bibfield  {title}
  {\enquote {\bibinfo {title} {Net charge per residue modulates conformational
  ensembles of intrinsically disordered proteins},}\ }\href@noop {} {\bibfield
  {journal} {\bibinfo  {journal} {Proceedings of the National Academy of
  Sciences}\ }\textbf {\bibinfo {volume} {107}},\ \bibinfo {pages} {8183--8188}
  (\bibinfo {year} {2010})}\BibitemShut {NoStop}%
\bibitem [{\citenamefont {Uversky}(2002)}]{uversky2002does}%
  \BibitemOpen
  \bibfield  {author} {\bibinfo {author} {\bibfnamefont {V.~N.}\ \bibnamefont
  {Uversky}},\ }\bibfield  {title} {\enquote {\bibinfo {title} {What does it
  mean to be natively unfolded?}}\ }\href@noop {} {\bibfield  {journal}
  {\bibinfo  {journal} {European journal of biochemistry}\ }\textbf {\bibinfo
  {volume} {269}},\ \bibinfo {pages} {2--12} (\bibinfo {year}
  {2002})}\BibitemShut {NoStop}%
\bibitem [{\citenamefont {Xie}\ \emph {et~al.}(2007)\citenamefont {Xie},
  \citenamefont {Vucetic}, \citenamefont {Iakoucheva}, \citenamefont
  {Oldfield}, \citenamefont {Dunker}, \citenamefont {Uversky},\ and\
  \citenamefont {Obradovic}}]{xie2007functional}%
  \BibitemOpen
  \bibfield  {author} {\bibinfo {author} {\bibfnamefont {H.}~\bibnamefont
  {Xie}}, \bibinfo {author} {\bibfnamefont {S.}~\bibnamefont {Vucetic}},
  \bibinfo {author} {\bibfnamefont {L.~M.}\ \bibnamefont {Iakoucheva}},
  \bibinfo {author} {\bibfnamefont {C.~J.}\ \bibnamefont {Oldfield}}, \bibinfo
  {author} {\bibfnamefont {A.~K.}\ \bibnamefont {Dunker}}, \bibinfo {author}
  {\bibfnamefont {V.~N.}\ \bibnamefont {Uversky}},\ and\ \bibinfo {author}
  {\bibfnamefont {Z.}~\bibnamefont {Obradovic}},\ }\bibfield  {title} {\enquote
  {\bibinfo {title} {Functional anthology of intrinsic disorder. 1. biological
  processes and functions of proteins with long disordered regions},}\
  }\href@noop {} {\bibfield  {journal} {\bibinfo  {journal} {Journal of
  proteome research}\ }\textbf {\bibinfo {volume} {6}},\ \bibinfo {pages}
  {1882--1898} (\bibinfo {year} {2007})}\BibitemShut {NoStop}%
\bibitem [{\citenamefont {Tompa}(2012)}]{tompa2012intrinsically}%
  \BibitemOpen
  \bibfield  {author} {\bibinfo {author} {\bibfnamefont {P.}~\bibnamefont
  {Tompa}},\ }\bibfield  {title} {\enquote {\bibinfo {title} {Intrinsically
  disordered proteins: a 10-year recap},}\ }\href@noop {} {\bibfield  {journal}
  {\bibinfo  {journal} {Trends in biochemical sciences}\ }\textbf {\bibinfo
  {volume} {37}},\ \bibinfo {pages} {509--516} (\bibinfo {year}
  {2012})}\BibitemShut {NoStop}%
\bibitem [{\citenamefont {Samanta}, \citenamefont {Chakraborty},\ and\
  \citenamefont {Thirumalai}(2018)}]{samanta2018charge}%
  \BibitemOpen
  \bibfield  {author} {\bibinfo {author} {\bibfnamefont {H.~S.}\ \bibnamefont
  {Samanta}}, \bibinfo {author} {\bibfnamefont {D.}~\bibnamefont
  {Chakraborty}},\ and\ \bibinfo {author} {\bibfnamefont {D.}~\bibnamefont
  {Thirumalai}},\ }\bibfield  {title} {\enquote {\bibinfo {title} {Charge
  fluctuation effects on the shape of flexible polyampholytes with applications
  to intrinsically disordered proteins},}\ }\href@noop {} {\bibfield  {journal}
  {\bibinfo  {journal} {The Journal of chemical physics}\ }\textbf {\bibinfo
  {volume} {149}},\ \bibinfo {pages} {163323} (\bibinfo {year}
  {2018})}\BibitemShut {NoStop}%
\bibitem [{\citenamefont {Bianchi}\ \emph {et~al.}(2020)\citenamefont
  {Bianchi}, \citenamefont {Longhi}, \citenamefont {Grandori},\ and\
  \citenamefont {Brocca}}]{bianchi2020relevance}%
  \BibitemOpen
  \bibfield  {author} {\bibinfo {author} {\bibfnamefont {G.}~\bibnamefont
  {Bianchi}}, \bibinfo {author} {\bibfnamefont {S.}~\bibnamefont {Longhi}},
  \bibinfo {author} {\bibfnamefont {R.}~\bibnamefont {Grandori}},\ and\
  \bibinfo {author} {\bibfnamefont {S.}~\bibnamefont {Brocca}},\ }\bibfield
  {title} {\enquote {\bibinfo {title} {Relevance of electrostatic charges in
  compactness, aggregation, and phase separation of intrinsically disordered
  proteins},}\ }\href@noop {} {\bibfield  {journal} {\bibinfo  {journal}
  {International Journal of Molecular Sciences}\ }\textbf {\bibinfo {volume}
  {21}},\ \bibinfo {pages} {6208} (\bibinfo {year} {2020})}\BibitemShut
  {NoStop}%
\bibitem [{\citenamefont {Danielsen}\ \emph {et~al.}(2019)\citenamefont
  {Danielsen}, \citenamefont {McCarty}, \citenamefont {Shea}, \citenamefont
  {Delaney},\ and\ \citenamefont {Fredrickson}}]{danielsen2019molecular}%
  \BibitemOpen
  \bibfield  {author} {\bibinfo {author} {\bibfnamefont {S.~P.}\ \bibnamefont
  {Danielsen}}, \bibinfo {author} {\bibfnamefont {J.}~\bibnamefont {McCarty}},
  \bibinfo {author} {\bibfnamefont {J.-E.}\ \bibnamefont {Shea}}, \bibinfo
  {author} {\bibfnamefont {K.~T.}\ \bibnamefont {Delaney}},\ and\ \bibinfo
  {author} {\bibfnamefont {G.~H.}\ \bibnamefont {Fredrickson}},\ }\bibfield
  {title} {\enquote {\bibinfo {title} {Molecular design of self-coacervation
  phenomena in block polyampholytes},}\ }\href@noop {} {\bibfield  {journal}
  {\bibinfo  {journal} {Proceedings of the National Academy of Sciences}\
  }\textbf {\bibinfo {volume} {116}},\ \bibinfo {pages} {8224--8232} (\bibinfo
  {year} {2019})}\BibitemShut {NoStop}%
\bibitem [{\citenamefont {M{\"u}ller-Sp{\"a}th}\ \emph
  {et~al.}(2010)\citenamefont {M{\"u}ller-Sp{\"a}th}, \citenamefont {Soranno},
  \citenamefont {Hirschfeld}, \citenamefont {Hofmann}, \citenamefont
  {R{\"u}egger}, \citenamefont {Reymond}, \citenamefont {Nettels},\ and\
  \citenamefont {Schuler}}]{muller2010charge}%
  \BibitemOpen
  \bibfield  {author} {\bibinfo {author} {\bibfnamefont {S.}~\bibnamefont
  {M{\"u}ller-Sp{\"a}th}}, \bibinfo {author} {\bibfnamefont {A.}~\bibnamefont
  {Soranno}}, \bibinfo {author} {\bibfnamefont {V.}~\bibnamefont {Hirschfeld}},
  \bibinfo {author} {\bibfnamefont {H.}~\bibnamefont {Hofmann}}, \bibinfo
  {author} {\bibfnamefont {S.}~\bibnamefont {R{\"u}egger}}, \bibinfo {author}
  {\bibfnamefont {L.}~\bibnamefont {Reymond}}, \bibinfo {author} {\bibfnamefont
  {D.}~\bibnamefont {Nettels}},\ and\ \bibinfo {author} {\bibfnamefont
  {B.}~\bibnamefont {Schuler}},\ }\bibfield  {title} {\enquote {\bibinfo
  {title} {Charge interactions can dominate the dimensions of intrinsically
  disordered proteins},}\ }\href@noop {} {\bibfield  {journal} {\bibinfo
  {journal} {Proceedings of the National Academy of Sciences}\ }\textbf
  {\bibinfo {volume} {107}},\ \bibinfo {pages} {14609--14614} (\bibinfo {year}
  {2010})}\BibitemShut {NoStop}%
\bibitem [{\citenamefont {Sundaravadivelu~Devarajan}\ \emph
  {et~al.}(2024)\citenamefont {Sundaravadivelu~Devarajan}, \citenamefont
  {Wang}, \citenamefont {Sza{\l}a-Mendyk}, \citenamefont {Rekhi}, \citenamefont
  {Nikoubashman}, \citenamefont {Kim},\ and\ \citenamefont
  {Mittal}}]{sundaravadivelu2024sequence}%
  \BibitemOpen
  \bibfield  {author} {\bibinfo {author} {\bibfnamefont {D.}~\bibnamefont
  {Sundaravadivelu~Devarajan}}, \bibinfo {author} {\bibfnamefont
  {J.}~\bibnamefont {Wang}}, \bibinfo {author} {\bibfnamefont {B.}~\bibnamefont
  {Sza{\l}a-Mendyk}}, \bibinfo {author} {\bibfnamefont {S.}~\bibnamefont
  {Rekhi}}, \bibinfo {author} {\bibfnamefont {A.}~\bibnamefont {Nikoubashman}},
  \bibinfo {author} {\bibfnamefont {Y.~C.}\ \bibnamefont {Kim}},\ and\ \bibinfo
  {author} {\bibfnamefont {J.}~\bibnamefont {Mittal}},\ }\bibfield  {title}
  {\enquote {\bibinfo {title} {Sequence-dependent material properties of
  biomolecular condensates and their relation to dilute phase conformations},}\
  }\href@noop {} {\bibfield  {journal} {\bibinfo  {journal} {Nature
  Communications}\ }\textbf {\bibinfo {volume} {15}},\ \bibinfo {pages} {1912}
  (\bibinfo {year} {2024})}\BibitemShut {NoStop}%
\bibitem [{\citenamefont {Gohy}\ \emph {et~al.}(2000)\citenamefont {Gohy},
  \citenamefont {Creutz}, \citenamefont {Garcia}, \citenamefont {Mahltig},
  \citenamefont {Stamm},\ and\ \citenamefont
  {J{\'e}r{\^o}me}}]{gohy2000aggregates}%
  \BibitemOpen
  \bibfield  {author} {\bibinfo {author} {\bibfnamefont {J.-F.}\ \bibnamefont
  {Gohy}}, \bibinfo {author} {\bibfnamefont {S.}~\bibnamefont {Creutz}},
  \bibinfo {author} {\bibfnamefont {M.}~\bibnamefont {Garcia}}, \bibinfo
  {author} {\bibfnamefont {B.}~\bibnamefont {Mahltig}}, \bibinfo {author}
  {\bibfnamefont {M.}~\bibnamefont {Stamm}},\ and\ \bibinfo {author}
  {\bibfnamefont {R.}~\bibnamefont {J{\'e}r{\^o}me}},\ }\bibfield  {title}
  {\enquote {\bibinfo {title} {Aggregates formed by amphoteric diblock
  copolymers in water},}\ }\href@noop {} {\bibfield  {journal} {\bibinfo
  {journal} {Macromolecules}\ }\textbf {\bibinfo {volume} {33}},\ \bibinfo
  {pages} {6378--6387} (\bibinfo {year} {2000})}\BibitemShut {NoStop}%
\bibitem [{\citenamefont {Devarajan}\ \emph {et~al.}(2022)\citenamefont
  {Devarajan}, \citenamefont {Rekhi}, \citenamefont {Nikoubashman},
  \citenamefont {Kim}, \citenamefont {Howard},\ and\ \citenamefont
  {Mittal}}]{devarajan2022effect}%
  \BibitemOpen
  \bibfield  {author} {\bibinfo {author} {\bibfnamefont {D.~S.}\ \bibnamefont
  {Devarajan}}, \bibinfo {author} {\bibfnamefont {S.}~\bibnamefont {Rekhi}},
  \bibinfo {author} {\bibfnamefont {A.}~\bibnamefont {Nikoubashman}}, \bibinfo
  {author} {\bibfnamefont {Y.~C.}\ \bibnamefont {Kim}}, \bibinfo {author}
  {\bibfnamefont {M.~P.}\ \bibnamefont {Howard}},\ and\ \bibinfo {author}
  {\bibfnamefont {J.}~\bibnamefont {Mittal}},\ }\bibfield  {title} {\enquote
  {\bibinfo {title} {Effect of charge distribution on the dynamics of
  polyampholytic disordered proteins},}\ }\href@noop {} {\bibfield  {journal}
  {\bibinfo  {journal} {Macromolecules}\ }\textbf {\bibinfo {volume} {55}},\
  \bibinfo {pages} {8987--8997} (\bibinfo {year} {2022})}\BibitemShut {NoStop}%
\bibitem [{\citenamefont {Goloub}, \citenamefont {De~Keizer},\ and\
  \citenamefont {Cohen~Stuart}(1999)}]{goloub1999association}%
  \BibitemOpen
  \bibfield  {author} {\bibinfo {author} {\bibfnamefont {T.}~\bibnamefont
  {Goloub}}, \bibinfo {author} {\bibfnamefont {A.}~\bibnamefont {De~Keizer}},\
  and\ \bibinfo {author} {\bibfnamefont {M.}~\bibnamefont {Cohen~Stuart}},\
  }\bibfield  {title} {\enquote {\bibinfo {title} {Association behavior of
  ampholytic diblock copolymers},}\ }\href@noop {} {\bibfield  {journal}
  {\bibinfo  {journal} {Macromolecules}\ }\textbf {\bibinfo {volume} {32}},\
  \bibinfo {pages} {8441--8446} (\bibinfo {year} {1999})}\BibitemShut {NoStop}%
\bibitem [{\citenamefont {Radhakrishnan}\ and\ \citenamefont
  {Singh}(2021)}]{radhakrishnan2021collapse}%
  \BibitemOpen
  \bibfield  {author} {\bibinfo {author} {\bibfnamefont {K.}~\bibnamefont
  {Radhakrishnan}}\ and\ \bibinfo {author} {\bibfnamefont {S.~P.}\ \bibnamefont
  {Singh}},\ }\bibfield  {title} {\enquote {\bibinfo {title} {Collapse of a
  confined polyelectrolyte chain under an ac electric field},}\ }\href@noop {}
  {\bibfield  {journal} {\bibinfo  {journal} {Macromolecules}\ }\textbf
  {\bibinfo {volume} {54}},\ \bibinfo {pages} {7998--8007} (\bibinfo {year}
  {2021})}\BibitemShut {NoStop}%
\bibitem [{\citenamefont {Imbert}\ \emph {et~al.}(1999)\citenamefont {Imbert},
  \citenamefont {Victor}, \citenamefont {Tsunekawa},\ and\ \citenamefont
  {Hiwatari}}]{imbert1999conformational}%
  \BibitemOpen
  \bibfield  {author} {\bibinfo {author} {\bibfnamefont {J.}~\bibnamefont
  {Imbert}}, \bibinfo {author} {\bibfnamefont {J.}~\bibnamefont {Victor}},
  \bibinfo {author} {\bibfnamefont {N.}~\bibnamefont {Tsunekawa}},\ and\
  \bibinfo {author} {\bibfnamefont {Y.}~\bibnamefont {Hiwatari}},\ }\bibfield
  {title} {\enquote {\bibinfo {title} {Conformational transitions of a diblock
  polyampholyte in 2 and 3 dimensions},}\ }\href@noop {} {\bibfield  {journal}
  {\bibinfo  {journal} {Physics Letters A}\ }\textbf {\bibinfo {volume}
  {258}},\ \bibinfo {pages} {92--98} (\bibinfo {year} {1999})}\BibitemShut
  {NoStop}%
\bibitem [{\citenamefont {Edwards}, \citenamefont {King},\ and\ \citenamefont
  {Pincus}(1980)}]{edwards1980phase}%
  \BibitemOpen
  \bibfield  {author} {\bibinfo {author} {\bibfnamefont {S.}~\bibnamefont
  {Edwards}}, \bibinfo {author} {\bibfnamefont {P.}~\bibnamefont {King}},\ and\
  \bibinfo {author} {\bibfnamefont {P.}~\bibnamefont {Pincus}},\ }\bibfield
  {title} {\enquote {\bibinfo {title} {Phase changes in polyampholytes},}\
  }\href@noop {} {\bibfield  {journal} {\bibinfo  {journal} {Ferroelectrics}\
  }\textbf {\bibinfo {volume} {30}},\ \bibinfo {pages} {3--6} (\bibinfo {year}
  {1980})}\BibitemShut {NoStop}%
\bibitem [{\citenamefont {Kantor}\ and\ \citenamefont
  {Kardar}(1991)}]{kantor1991polymers}%
  \BibitemOpen
  \bibfield  {author} {\bibinfo {author} {\bibfnamefont {Y.}~\bibnamefont
  {Kantor}}\ and\ \bibinfo {author} {\bibfnamefont {M.}~\bibnamefont
  {Kardar}},\ }\bibfield  {title} {\enquote {\bibinfo {title} {Polymers with
  random self-interactions},}\ }\href@noop {} {\bibfield  {journal} {\bibinfo
  {journal} {Europhysics Letters}\ }\textbf {\bibinfo {volume} {14}},\ \bibinfo
  {pages} {421} (\bibinfo {year} {1991})}\BibitemShut {NoStop}%
\bibitem [{\citenamefont {Dobrynin}\ and\ \citenamefont
  {Rubinstein}(1995)}]{dobrynin1995flory}%
  \BibitemOpen
  \bibfield  {author} {\bibinfo {author} {\bibfnamefont {A.~V.}\ \bibnamefont
  {Dobrynin}}\ and\ \bibinfo {author} {\bibfnamefont {M.}~\bibnamefont
  {Rubinstein}},\ }\bibfield  {title} {\enquote {\bibinfo {title} {Flory theory
  of a polyampholyte chain},}\ }\href@noop {} {\bibfield  {journal} {\bibinfo
  {journal} {Journal de Physique II}\ }\textbf {\bibinfo {volume} {5}},\
  \bibinfo {pages} {677--695} (\bibinfo {year} {1995})}\BibitemShut {NoStop}%
\bibitem [{\citenamefont {Jeon}\ and\ \citenamefont
  {Dobrynin}(2005)}]{jeon2005molecular}%
  \BibitemOpen
  \bibfield  {author} {\bibinfo {author} {\bibfnamefont {J.}~\bibnamefont
  {Jeon}}\ and\ \bibinfo {author} {\bibfnamefont {A.~V.}\ \bibnamefont
  {Dobrynin}},\ }\bibfield  {title} {\enquote {\bibinfo {title} {Molecular
  dynamics simulations of polyampholyte- polyelectrolyte complexes in
  solutions},}\ }\href@noop {} {\bibfield  {journal} {\bibinfo  {journal}
  {Macromolecules}\ }\textbf {\bibinfo {volume} {38}},\ \bibinfo {pages}
  {5300--5312} (\bibinfo {year} {2005})}\BibitemShut {NoStop}%
\bibitem [{\citenamefont {Sawle}\ and\ \citenamefont
  {Ghosh}(2015)}]{sawle2015theoretical}%
  \BibitemOpen
  \bibfield  {author} {\bibinfo {author} {\bibfnamefont {L.}~\bibnamefont
  {Sawle}}\ and\ \bibinfo {author} {\bibfnamefont {K.}~\bibnamefont {Ghosh}},\
  }\bibfield  {title} {\enquote {\bibinfo {title} {A theoretical method to
  compute sequence dependent configurational properties in charged polymers and
  proteins},}\ }\href@noop {} {\bibfield  {journal} {\bibinfo  {journal} {The
  Journal of chemical physics}\ }\textbf {\bibinfo {volume} {143}} (\bibinfo
  {year} {2015})}\BibitemShut {NoStop}%
\bibitem [{\citenamefont {Baumketner}\ \emph {et~al.}(2001)\citenamefont
  {Baumketner}, \citenamefont {Shimizu}, \citenamefont {Isobe},\ and\
  \citenamefont {Hiwatari}}]{baumketner2001helix}%
  \BibitemOpen
  \bibfield  {author} {\bibinfo {author} {\bibfnamefont {A.}~\bibnamefont
  {Baumketner}}, \bibinfo {author} {\bibfnamefont {H.}~\bibnamefont {Shimizu}},
  \bibinfo {author} {\bibfnamefont {M.}~\bibnamefont {Isobe}},\ and\ \bibinfo
  {author} {\bibfnamefont {Y.}~\bibnamefont {Hiwatari}},\ }\bibfield  {title}
  {\enquote {\bibinfo {title} {Helix transition in di-block polyampholyte},}\
  }\href@noop {} {\bibfield  {journal} {\bibinfo  {journal} {Journal of
  Physics: Condensed Matter}\ }\textbf {\bibinfo {volume} {13}},\ \bibinfo
  {pages} {10279} (\bibinfo {year} {2001})}\BibitemShut {NoStop}%
\bibitem [{\citenamefont {Baratlo}\ and\ \citenamefont
  {Fazli}(2010)}]{baratlo2010brushes}%
  \BibitemOpen
  \bibfield  {author} {\bibinfo {author} {\bibfnamefont {M.}~\bibnamefont
  {Baratlo}}\ and\ \bibinfo {author} {\bibfnamefont {H.}~\bibnamefont
  {Fazli}},\ }\bibfield  {title} {\enquote {\bibinfo {title} {Brushes of
  flexible, semiflexible, and rodlike diblock polyampholytes: Molecular
  dynamics simulation and scaling analysis},}\ }\href@noop {} {\bibfield
  {journal} {\bibinfo  {journal} {Physical Review E}\ }\textbf {\bibinfo
  {volume} {81}},\ \bibinfo {pages} {011801} (\bibinfo {year}
  {2010})}\BibitemShut {NoStop}%
\bibitem [{\citenamefont {Akinchina}\ and\ \citenamefont
  {Linse}(2007)}]{akinchina2007diblock}%
  \BibitemOpen
  \bibfield  {author} {\bibinfo {author} {\bibfnamefont {A.}~\bibnamefont
  {Akinchina}}\ and\ \bibinfo {author} {\bibfnamefont {P.}~\bibnamefont
  {Linse}},\ }\bibfield  {title} {\enquote {\bibinfo {title} {Diblock
  polyampholytes grafted onto spherical particles: effect of stiffness, charge
  density, and grafting density},}\ }\href@noop {} {\bibfield  {journal}
  {\bibinfo  {journal} {Langmuir}\ }\textbf {\bibinfo {volume} {23}},\ \bibinfo
  {pages} {1465--1472} (\bibinfo {year} {2007})}\BibitemShut {NoStop}%
\bibitem [{\citenamefont {Zierenberg}, \citenamefont {Marenz},\ and\
  \citenamefont {Janke}(2016)}]{zierenberg2016dilute}%
  \BibitemOpen
  \bibfield  {author} {\bibinfo {author} {\bibfnamefont {J.}~\bibnamefont
  {Zierenberg}}, \bibinfo {author} {\bibfnamefont {M.}~\bibnamefont {Marenz}},\
  and\ \bibinfo {author} {\bibfnamefont {W.}~\bibnamefont {Janke}},\ }\bibfield
   {title} {\enquote {\bibinfo {title} {Dilute semiflexible polymers with
  attraction: Collapse, folding and aggregation},}\ }\href@noop {} {\bibfield
  {journal} {\bibinfo  {journal} {Polymers}\ }\textbf {\bibinfo {volume} {8}},\
  \bibinfo {pages} {333} (\bibinfo {year} {2016})}\BibitemShut {NoStop}%
\bibitem [{\citenamefont {De~Gennes}\ and\ \citenamefont
  {Gennes}(1979)}]{de1979scaling}%
  \BibitemOpen
  \bibfield  {author} {\bibinfo {author} {\bibfnamefont {P.-G.}\ \bibnamefont
  {De~Gennes}}\ and\ \bibinfo {author} {\bibfnamefont {P.-G.}\ \bibnamefont
  {Gennes}},\ }\href@noop {} {\emph {\bibinfo {title} {Scaling concepts in
  polymer physics}}}\ (\bibinfo  {publisher} {Cornell university press},\
  \bibinfo {year} {1979})\BibitemShut {NoStop}%
\bibitem [{\citenamefont {Seaton}\ \emph {et~al.}(2013)\citenamefont {Seaton},
  \citenamefont {Schnabel}, \citenamefont {Landau},\ and\ \citenamefont
  {Bachmann}}]{seaton2013flexible}%
  \BibitemOpen
  \bibfield  {author} {\bibinfo {author} {\bibfnamefont {D.~T.}\ \bibnamefont
  {Seaton}}, \bibinfo {author} {\bibfnamefont {S.}~\bibnamefont {Schnabel}},
  \bibinfo {author} {\bibfnamefont {D.~P.}\ \bibnamefont {Landau}},\ and\
  \bibinfo {author} {\bibfnamefont {M.}~\bibnamefont {Bachmann}},\ }\bibfield
  {title} {\enquote {\bibinfo {title} {From flexible to stiff: Systematic
  analysis of structural phases for single semiflexible polymers},}\
  }\href@noop {} {\bibfield  {journal} {\bibinfo  {journal} {Physical review
  letters}\ }\textbf {\bibinfo {volume} {110}},\ \bibinfo {pages} {028103}
  (\bibinfo {year} {2013})}\BibitemShut {NoStop}%
\bibitem [{\citenamefont {Marenz}\ and\ \citenamefont
  {Janke}(2016)}]{marenz2016knots}%
  \BibitemOpen
  \bibfield  {author} {\bibinfo {author} {\bibfnamefont {M.}~\bibnamefont
  {Marenz}}\ and\ \bibinfo {author} {\bibfnamefont {W.}~\bibnamefont {Janke}},\
  }\bibfield  {title} {\enquote {\bibinfo {title} {Knots as a topological order
  parameter for semiflexible polymers},}\ }\href@noop {} {\bibfield  {journal}
  {\bibinfo  {journal} {Physical review letters}\ }\textbf {\bibinfo {volume}
  {116}},\ \bibinfo {pages} {128301} (\bibinfo {year} {2016})}\BibitemShut
  {NoStop}%
\bibitem [{\citenamefont {Huang}\ \emph {et~al.}(2016)\citenamefont {Huang},
  \citenamefont {Huang}, \citenamefont {Lei},\ and\ \citenamefont
  {Larson}}]{huang2016simple}%
  \BibitemOpen
  \bibfield  {author} {\bibinfo {author} {\bibfnamefont {W.}~\bibnamefont
  {Huang}}, \bibinfo {author} {\bibfnamefont {M.}~\bibnamefont {Huang}},
  \bibinfo {author} {\bibfnamefont {Q.}~\bibnamefont {Lei}},\ and\ \bibinfo
  {author} {\bibfnamefont {R.~G.}\ \bibnamefont {Larson}},\ }\bibfield  {title}
  {\enquote {\bibinfo {title} {A simple analytical model for predicting the
  collapsed state of self-attractive semiflexible polymers},}\ }\href@noop {}
  {\bibfield  {journal} {\bibinfo  {journal} {Polymers}\ }\textbf {\bibinfo
  {volume} {8}},\ \bibinfo {pages} {264} (\bibinfo {year} {2016})}\BibitemShut
  {NoStop}%
\bibitem [{\citenamefont {Aierken}\ and\ \citenamefont
  {Bachmann}(2023{\natexlab{a}})}]{aierken2023impact}%
  \BibitemOpen
  \bibfield  {author} {\bibinfo {author} {\bibfnamefont {D.}~\bibnamefont
  {Aierken}}\ and\ \bibinfo {author} {\bibfnamefont {M.}~\bibnamefont
  {Bachmann}},\ }\bibfield  {title} {\enquote {\bibinfo {title} {Impact of
  bending stiffness on ground-state conformations for semiflexible polymers},}\
  }\href@noop {} {\bibfield  {journal} {\bibinfo  {journal} {The Journal of
  Chemical Physics}\ }\textbf {\bibinfo {volume} {158}} (\bibinfo {year}
  {2023}{\natexlab{a}})}\BibitemShut {NoStop}%
\bibitem [{\citenamefont {Gordievskaya}, \citenamefont {Budkov},\ and\
  \citenamefont {Kramarenko}(2018)}]{gordievskaya2018interplay}%
  \BibitemOpen
  \bibfield  {author} {\bibinfo {author} {\bibfnamefont {Y.~D.}\ \bibnamefont
  {Gordievskaya}}, \bibinfo {author} {\bibfnamefont {Y.~A.}\ \bibnamefont
  {Budkov}},\ and\ \bibinfo {author} {\bibfnamefont {E.~Y.}\ \bibnamefont
  {Kramarenko}},\ }\bibfield  {title} {\enquote {\bibinfo {title} {An interplay
  of electrostatic and excluded volume interactions in the conformational
  behavior of a dipolar chain: theory and computer simulations},}\ }\href@noop
  {} {\bibfield  {journal} {\bibinfo  {journal} {Soft Matter}\ }\textbf
  {\bibinfo {volume} {14}},\ \bibinfo {pages} {3232--3235} (\bibinfo {year}
  {2018})}\BibitemShut {NoStop}%
\bibitem [{\citenamefont {Hsieh}, \citenamefont {Jain},\ and\ \citenamefont
  {Larson}(2006)}]{hsieh2006brownian}%
  \BibitemOpen
  \bibfield  {author} {\bibinfo {author} {\bibfnamefont {C.-C.}\ \bibnamefont
  {Hsieh}}, \bibinfo {author} {\bibfnamefont {S.}~\bibnamefont {Jain}},\ and\
  \bibinfo {author} {\bibfnamefont {R.~G.}\ \bibnamefont {Larson}},\ }\bibfield
   {title} {\enquote {\bibinfo {title} {Brownian dynamics simulations with
  stiff finitely extensible nonlinear elastic-fraenkel springs as
  approximations to rods in bead-rod models},}\ }\href@noop {} {\bibfield
  {journal} {\bibinfo  {journal} {The Journal of chemical physics}\ }\textbf
  {\bibinfo {volume} {124}} (\bibinfo {year} {2006})}\BibitemShut {NoStop}%
\bibitem [{LAM()}]{LAMMPS}%
  \BibitemOpen
  \href@noop {} {\emph {\bibinfo {title} {LAMMPS Molecular Dynamics
  Simulator}}}\ (\bibinfo  {publisher} {Sandia National Laboratories},\
  \bibinfo {address} {http://lammps.sandia.gov})\BibitemShut {NoStop}%
\bibitem [{\citenamefont {Rubinstein}\ and\ \citenamefont
  {Colby}(2003)}]{Rubi:2003}%
  \BibitemOpen
  \bibfield  {author} {\bibinfo {author} {\bibfnamefont {M.}~\bibnamefont
  {Rubinstein}}\ and\ \bibinfo {author} {\bibfnamefont {R.~H.}\ \bibnamefont
  {Colby}},\ }\href@noop {} {\emph {\bibinfo {title} {Polymer physics}}},\
  Vol.~\bibinfo {volume} {23}\ (\bibinfo  {publisher} {Oxford University Press
  New York},\ \bibinfo {year} {2003})\BibitemShut {NoStop}%
\bibitem [{\citenamefont {Doi}\ and\ \citenamefont {Edwards}(1986)}]{doi:86}%
  \BibitemOpen
  \bibfield  {author} {\bibinfo {author} {\bibfnamefont {M.}~\bibnamefont
  {Doi}}\ and\ \bibinfo {author} {\bibfnamefont {S.~F.}\ \bibnamefont
  {Edwards}},\ }\href@noop {} {\emph {\bibinfo {title} {The Theory of Polymer
  Dynamics}}}\ (\bibinfo  {publisher} {Oxford University press},\ \bibinfo
  {address} {New York},\ \bibinfo {year} {1986})\BibitemShut {NoStop}%
\bibitem [{\citenamefont {Muthukumar}(2011)}]{Muthu_Book_2011}%
  \BibitemOpen
  \bibfield  {author} {\bibinfo {author} {\bibfnamefont {M.}~\bibnamefont
  {Muthukumar}},\ }\href@noop {} {\emph {\bibinfo {title} {Polymer
  translocation}}}\ (\bibinfo  {publisher} {Taylor \& Francis US},\ \bibinfo
  {year} {2011})\BibitemShut {NoStop}%
\bibitem [{\citenamefont {de~Gennes}(1979)}]{Gennes_SCP_1979}%
  \BibitemOpen
  \bibfield  {author} {\bibinfo {author} {\bibfnamefont {P.~G.}\ \bibnamefont
  {de~Gennes}},\ }\href@noop {} {\emph {\bibinfo {title} {Scaling concepts in
  polymer physics}}}\ (\bibinfo  {publisher} {Cornell University Press},\
  \bibinfo {address} {Ithaca},\ \bibinfo {year} {1979})\BibitemShut {NoStop}%
\bibitem [{\citenamefont {Deserno}\ and\ \citenamefont
  {Holm}(1998{\natexlab{a}})}]{deserno1998mesh_1}%
  \BibitemOpen
  \bibfield  {author} {\bibinfo {author} {\bibfnamefont {M.}~\bibnamefont
  {Deserno}}\ and\ \bibinfo {author} {\bibfnamefont {C.}~\bibnamefont {Holm}},\
  }\bibfield  {title} {\enquote {\bibinfo {title} {How to mesh up ewald sums.
  i. a theoretical and numerical comparison of various particle mesh
  routines},}\ }\href@noop {} {\bibfield  {journal} {\bibinfo  {journal} {The
  Journal of chemical physics}\ }\textbf {\bibinfo {volume} {109}},\ \bibinfo
  {pages} {7678--7693} (\bibinfo {year} {1998}{\natexlab{a}})}\BibitemShut
  {NoStop}%
\bibitem [{\citenamefont {Deserno}\ and\ \citenamefont
  {Holm}(1998{\natexlab{b}})}]{deserno1998mesh_2}%
  \BibitemOpen
  \bibfield  {author} {\bibinfo {author} {\bibfnamefont {M.}~\bibnamefont
  {Deserno}}\ and\ \bibinfo {author} {\bibfnamefont {C.}~\bibnamefont {Holm}},\
  }\bibfield  {title} {\enquote {\bibinfo {title} {How to mesh up ewald sums.
  ii. an accurate error estimate for the particle--particle--particle-mesh
  algorithm},}\ }\href@noop {} {\bibfield  {journal} {\bibinfo  {journal} {The
  Journal of Chemical Physics}\ }\textbf {\bibinfo {volume} {109}},\ \bibinfo
  {pages} {7694--7701} (\bibinfo {year} {1998}{\natexlab{b}})}\BibitemShut
  {NoStop}%
\bibitem [{\citenamefont {Allen}\ and\ \citenamefont
  {Tildesley}(1987)}]{allen87}%
  \BibitemOpen
  \bibfield  {author} {\bibinfo {author} {\bibfnamefont {M.~P.}\ \bibnamefont
  {Allen}}\ and\ \bibinfo {author} {\bibfnamefont {D.~J.}\ \bibnamefont
  {Tildesley}},\ }\href@noop {} {\emph {\bibinfo {title} {Computer Simulation
  of Liquids}}}\ (\bibinfo  {publisher} {Clarendon Press},\ \bibinfo {address}
  {Oxford},\ \bibinfo {year} {1987})\BibitemShut {NoStop}%
\bibitem [{\citenamefont {Huihui}, \citenamefont {Firman},\ and\ \citenamefont
  {Ghosh}(2018)}]{huihui2018modulating}%
  \BibitemOpen
  \bibfield  {author} {\bibinfo {author} {\bibfnamefont {J.}~\bibnamefont
  {Huihui}}, \bibinfo {author} {\bibfnamefont {T.}~\bibnamefont {Firman}},\
  and\ \bibinfo {author} {\bibfnamefont {K.}~\bibnamefont {Ghosh}},\ }\bibfield
   {title} {\enquote {\bibinfo {title} {Modulating charge patterning and ionic
  strength as a strategy to induce conformational changes in intrinsically
  disordered proteins},}\ }\href@noop {} {\bibfield  {journal} {\bibinfo
  {journal} {The Journal of Chemical Physics}\ }\textbf {\bibinfo {volume}
  {149}},\ \bibinfo {pages} {085101} (\bibinfo {year} {2018})}\BibitemShut
  {NoStop}%
\bibitem [{\citenamefont {Srivastava}\ and\ \citenamefont
  {Muthukumar}(1996)}]{srivastava1996sequence}%
  \BibitemOpen
  \bibfield  {author} {\bibinfo {author} {\bibfnamefont {D.}~\bibnamefont
  {Srivastava}}\ and\ \bibinfo {author} {\bibfnamefont {M.}~\bibnamefont
  {Muthukumar}},\ }\bibfield  {title} {\enquote {\bibinfo {title} {Sequence
  dependence of conformations of polyampholytes},}\ }\href@noop {} {\bibfield
  {journal} {\bibinfo  {journal} {Macromolecules}\ }\textbf {\bibinfo {volume}
  {29}},\ \bibinfo {pages} {2324--2326} (\bibinfo {year} {1996})}\BibitemShut
  {NoStop}%
\bibitem [{\citenamefont {Dinic}, \citenamefont {Marciel},\ and\ \citenamefont
  {Tirrell}(2021)}]{dinic2021polyampholyte}%
  \BibitemOpen
  \bibfield  {author} {\bibinfo {author} {\bibfnamefont {J.}~\bibnamefont
  {Dinic}}, \bibinfo {author} {\bibfnamefont {A.~B.}\ \bibnamefont {Marciel}},\
  and\ \bibinfo {author} {\bibfnamefont {M.~V.}\ \bibnamefont {Tirrell}},\
  }\bibfield  {title} {\enquote {\bibinfo {title} {Polyampholyte physics:
  Liquid--liquid phase separation and biological condensates},}\ }\href@noop {}
  {\bibfield  {journal} {\bibinfo  {journal} {Current opinion in colloid \&
  interface science}\ }\textbf {\bibinfo {volume} {54}},\ \bibinfo {pages}
  {101457} (\bibinfo {year} {2021})}\BibitemShut {NoStop}%
\bibitem [{\citenamefont {Gompper}\ \emph {et~al.}(2009)\citenamefont
  {Gompper}, \citenamefont {Ihle}, \citenamefont {Kroll},\ and\ \citenamefont
  {Winkler}}]{Gompper_APS_2009}%
  \BibitemOpen
  \bibfield  {author} {\bibinfo {author} {\bibfnamefont {G.}~\bibnamefont
  {Gompper}}, \bibinfo {author} {\bibfnamefont {T.}~\bibnamefont {Ihle}},
  \bibinfo {author} {\bibfnamefont {D.~M.}\ \bibnamefont {Kroll}},\ and\
  \bibinfo {author} {\bibfnamefont {R.~G.}\ \bibnamefont {Winkler}},\
  }\bibfield  {title} {\enquote {\bibinfo {title} {Multi-particle collision
  dynamics: a particle-based mesoscale simulation approach to the hydrodynamics
  of complex fluids},}\ }\href@noop {} {\bibfield  {journal} {\bibinfo
  {journal} {Adv. Polym. Sci.}\ }\textbf {\bibinfo {volume} {221}},\ \bibinfo
  {pages} {1--87} (\bibinfo {year} {2009})}\BibitemShut {NoStop}%
\bibitem [{\citenamefont {Kapral}(2008)}]{Kapral_ACP_2008}%
  \BibitemOpen
  \bibfield  {author} {\bibinfo {author} {\bibfnamefont {R.}~\bibnamefont
  {Kapral}},\ }\bibfield  {title} {\enquote {\bibinfo {title} {Multiparticle
  collision dynamics: {S}imulation of complex systems on mesoscales},}\
  }\href@noop {} {\bibfield  {journal} {\bibinfo  {journal} {Adv. Chem. Phys.}\
  }\textbf {\bibinfo {volume} {140}},\ \bibinfo {pages} {89} (\bibinfo {year}
  {2008})}\BibitemShut {NoStop}%
\bibitem [{\citenamefont {Singh}\ and\ \citenamefont
  {Muthukumar}(2014)}]{Singh_2014_JCP}%
  \BibitemOpen
  \bibfield  {author} {\bibinfo {author} {\bibfnamefont {S.~P.}\ \bibnamefont
  {Singh}}\ and\ \bibinfo {author} {\bibfnamefont {M.}~\bibnamefont
  {Muthukumar}},\ }\bibfield  {title} {\enquote {\bibinfo {title}
  {Electrophoretic mobilities of counterions and a polymer in cylindrical
  pores},}\ }\href@noop {} {\bibfield  {journal} {\bibinfo  {journal} {The
  Journal of chemical physics}\ }\textbf {\bibinfo {volume} {141}},\ \bibinfo
  {pages} {09B610\_1} (\bibinfo {year} {2014})}\BibitemShut {NoStop}%
\bibitem [{\citenamefont {Mukherji}\ \emph {et~al.}(2017)\citenamefont
  {Mukherji}, \citenamefont {Marques}, \citenamefont {Stuehn},\ and\
  \citenamefont {Kremer}}]{mukherji2017depleted}%
  \BibitemOpen
  \bibfield  {author} {\bibinfo {author} {\bibfnamefont {D.}~\bibnamefont
  {Mukherji}}, \bibinfo {author} {\bibfnamefont {C.~M.}\ \bibnamefont
  {Marques}}, \bibinfo {author} {\bibfnamefont {T.}~\bibnamefont {Stuehn}},\
  and\ \bibinfo {author} {\bibfnamefont {K.}~\bibnamefont {Kremer}},\
  }\bibfield  {title} {\enquote {\bibinfo {title} {Depleted depletion drives
  polymer swelling in poor solvent mixtures},}\ }\href@noop {} {\bibfield
  {journal} {\bibinfo  {journal} {Nature communications}\ }\textbf {\bibinfo
  {volume} {8}},\ \bibinfo {pages} {1374} (\bibinfo {year} {2017})}\BibitemShut
  {NoStop}%
\bibitem [{\citenamefont {Flory}(1953)}]{flory1953principles}%
  \BibitemOpen
  \bibfield  {author} {\bibinfo {author} {\bibfnamefont {P.~J.}\ \bibnamefont
  {Flory}},\ }\href@noop {} {\emph {\bibinfo {title} {Principles of polymer
  chemistry}}}\ (\bibinfo  {publisher} {Cornell university press},\ \bibinfo
  {year} {1953})\BibitemShut {NoStop}%
\bibitem [{\citenamefont {Birshtein}\ and\ \citenamefont
  {Pryamitsyn}(1987)}]{birshtein1987theory}%
  \BibitemOpen
  \bibfield  {author} {\bibinfo {author} {\bibfnamefont {T.}~\bibnamefont
  {Birshtein}}\ and\ \bibinfo {author} {\bibfnamefont {V.}~\bibnamefont
  {Pryamitsyn}},\ }\bibfield  {title} {\enquote {\bibinfo {title} {Theory of
  the coil-globule transition},}\ }\href@noop {} {\bibfield  {journal}
  {\bibinfo  {journal} {Polymer Science USSR}\ }\textbf {\bibinfo {volume}
  {29}},\ \bibinfo {pages} {2039--2046} (\bibinfo {year} {1987})}\BibitemShut
  {NoStop}%
\bibitem [{\citenamefont {Singh}\ \emph {et~al.}(2012)\citenamefont {Singh},
  \citenamefont {Chatterji}, \citenamefont {Winkler},\ and\ \citenamefont
  {Gompper}}]{Singh_JPCM_2012}%
  \BibitemOpen
  \bibfield  {author} {\bibinfo {author} {\bibfnamefont {S.~P.}\ \bibnamefont
  {Singh}}, \bibinfo {author} {\bibfnamefont {A.}~\bibnamefont {Chatterji}},
  \bibinfo {author} {\bibfnamefont {R.~G.}\ \bibnamefont {Winkler}},\ and\
  \bibinfo {author} {\bibfnamefont {G.}~\bibnamefont {Gompper}},\ }\bibfield
  {title} {\enquote {\bibinfo {title} {Conformational and dynamical properties
  of ultra-soft colloids in semi-dilute solutions under shear flow},}\
  }\href@noop {} {\bibfield  {journal} {\bibinfo  {journal} {J. Phys.: Condens.
  Matter}\ }\textbf {\bibinfo {volume} {24}},\ \bibinfo {pages} {464103}
  (\bibinfo {year} {2012})}\BibitemShut {NoStop}%
\bibitem [{\citenamefont {Khalatur}(1980)}]{khalatur1980effect}%
  \BibitemOpen
  \bibfield  {author} {\bibinfo {author} {\bibfnamefont {P.}~\bibnamefont
  {Khalatur}},\ }\bibfield  {title} {\enquote {\bibinfo {title} {Effect of
  volume interactions on the shape of a polymer coil},}\ }\href@noop {}
  {\bibfield  {journal} {\bibinfo  {journal} {Polymer Science USSR}\ }\textbf
  {\bibinfo {volume} {22}},\ \bibinfo {pages} {2438--2448} (\bibinfo {year}
  {1980})}\BibitemShut {NoStop}%
\bibitem [{\citenamefont {Aierken}\ and\ \citenamefont
  {Bachmann}(2023{\natexlab{b}})}]{aierken2023stable}%
  \BibitemOpen
  \bibfield  {author} {\bibinfo {author} {\bibfnamefont {D.}~\bibnamefont
  {Aierken}}\ and\ \bibinfo {author} {\bibfnamefont {M.}~\bibnamefont
  {Bachmann}},\ }\bibfield  {title} {\enquote {\bibinfo {title} {Stable
  intermediate phase of secondary structures for semiflexible polymers},}\
  }\href@noop {} {\bibfield  {journal} {\bibinfo  {journal} {Physical Review
  E}\ }\textbf {\bibinfo {volume} {107}},\ \bibinfo {pages} {L032501} (\bibinfo
  {year} {2023}{\natexlab{b}})}\BibitemShut {NoStop}%
\bibitem [{\citenamefont {Pantawane}\ and\ \citenamefont
  {Gekle}(2022)}]{pantawane2022temperature}%
  \BibitemOpen
  \bibfield  {author} {\bibinfo {author} {\bibfnamefont {S.}~\bibnamefont
  {Pantawane}}\ and\ \bibinfo {author} {\bibfnamefont {S.}~\bibnamefont
  {Gekle}},\ }\bibfield  {title} {\enquote {\bibinfo {title}
  {Temperature-dependent conformation behavior of isolated poly
  (3-hexylthiopene) chains},}\ }\href@noop {} {\bibfield  {journal} {\bibinfo
  {journal} {Polymers}\ }\textbf {\bibinfo {volume} {14}},\ \bibinfo {pages}
  {550} (\bibinfo {year} {2022})}\BibitemShut {NoStop}%
\bibitem [{\citenamefont {Gordievskaya}\ and\ \citenamefont
  {Kramarenko}(2019)}]{gordievskaya2019conformational}%
  \BibitemOpen
  \bibfield  {author} {\bibinfo {author} {\bibfnamefont {Y.~D.}\ \bibnamefont
  {Gordievskaya}}\ and\ \bibinfo {author} {\bibfnamefont {E.~Y.}\ \bibnamefont
  {Kramarenko}},\ }\bibfield  {title} {\enquote {\bibinfo {title}
  {Conformational behavior of a semiflexible dipolar chain with a variable
  relative size of charged groups via molecular dynamics simulations},}\
  }\href@noop {} {\bibfield  {journal} {\bibinfo  {journal} {Soft matter}\
  }\textbf {\bibinfo {volume} {15}},\ \bibinfo {pages} {6073--6085} (\bibinfo
  {year} {2019})}\BibitemShut {NoStop}%
\end{thebibliography}

%

\end{document}


\preprint{APS/123-QED}

\title{Supplementary Material: Structural transitions of a Semi-Flexible Polyampholyte}
\thanks{A footnote to the article title}%

\author{Rakesh Palariya}
\email{rakesh20@iiserb.ac.in}

\author{ Sunil P. Singh  }
\email{spsingh@iiserb.ac.in}
\affiliation{Department of Physics,\\ Indian Institute Of Science Education and Research, \\Bhopal 462 066, Madhya Pradesh, India}

\date{\today}

\maketitle

\begin{figure}[t]
\includegraphics[width=\columnwidth]{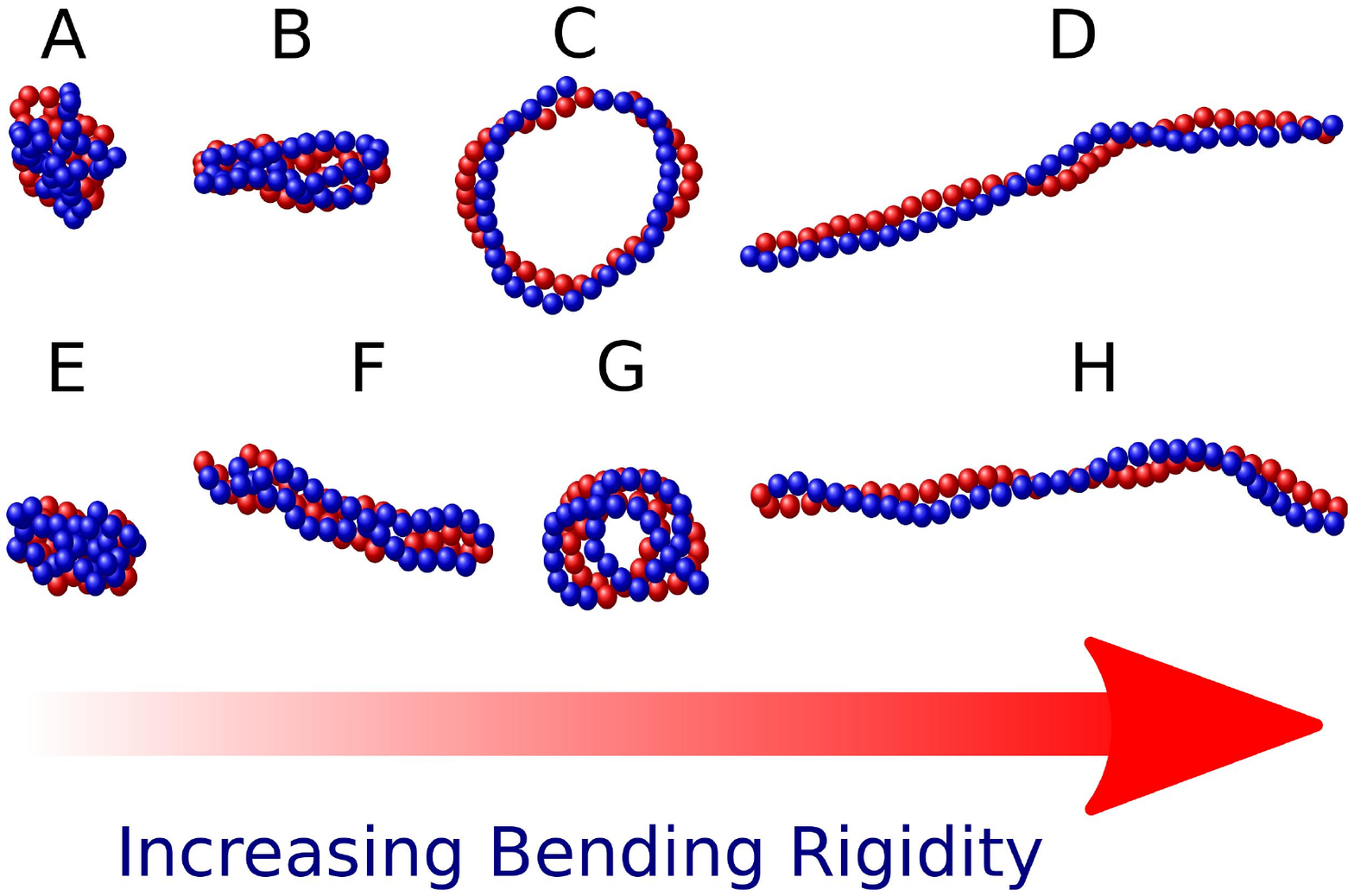}
\caption{The various conformational states of the PA chains upon increasing bending strengths  (left to right) for di-block and tetra-block PA chains for chain length $N=64$ and  $\Gamma_e=5$. Globular states (A and E), four-fold and hairpin states (B, F, D, and H), and circular/torus states (C and G). }
\label{Fig:structure}
\end{figure}

\subsection{ Multi-Particle Collision Dynamics} In this section, we present the hydrodynamic simulation model employed to investigate the dynamical behavior of the polyampholyte (PA) chain. The fluid surrounding the chain is simulated using a multi-particle collision dynamics (MPC) approach, an explicit solvent-based technique capable of accurately capturing thermal fluctuations and long-range hydrodynamic interactions.\cite{kapral2008multiparticle,Gompper2008MultiParticleCD} In this model, the solvent is represented by electrically neutral point particles.

The MPC approach involves two main dynamical steps. Firstly, each particle undergoes ballistic motion in the streaming step, resulting in updated positions and velocities at regular intervals (denoted as $h$). This process is described by the equation $r_i(t + h) = r_i(t) + h v_i(t)$, where $r_i(t)$ and $v_i(t)$ represent the position and velocity of particle $i$ at time $t$, respectively.

Secondly, the collision step occurs at every interval $h=0.1\tau$. The simulation box is divided into cubic cells of side length $a=\sigma$, and the solvent density is $\rho_s = 10m_s/\sigma^3$. Within each cell, the relative velocities of all particles w.r.t. the center of mass velocity are rotated around a random axis by an angle $\alpha$. The updated velocity of each particle is determined by a rotation matrix $\mathscr{R}(\alpha)$\cite{PhysRevE.66.036702}, as given by the equation
\begin{equation}
    v_i(t+h) = v_{cm}(t) + \mathscr{R}(\alpha)(v_i(t) - v_{cm}(t)).
    \label{Eq:vel}
\end{equation}
This collision protocol ensures the conservation of mass, energy, and momentum, thereby leading to long-range hydrodynamic interactions. The rotation angle here is considered to be $\alpha=130^o$. These parameters result in the solvent viscosity $\eta_s=8.7$ in dimensionless units. 

The velocities of the chain's monomers are included in every collision step to couple the PA chain with the solvent\cite{kapral2008multiparticle,Gompper2008MultiParticleCD,huang2010cell}. Additionally, a random grid shift is performed before the collision step to maintain Galilean invariance\cite{PhysRevE.63.020201}.

\begin{figure}[t]
\includegraphics[width=\linewidth]{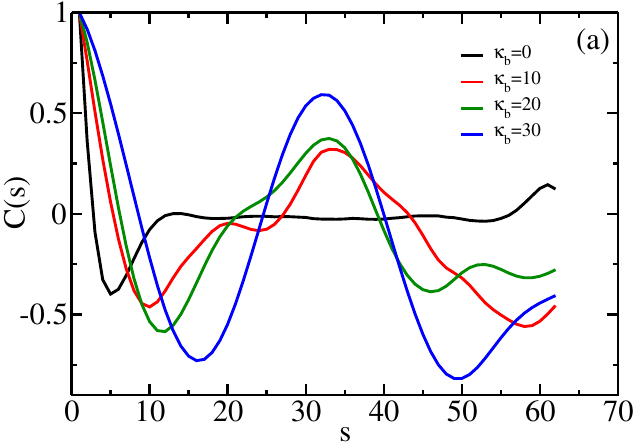}
\includegraphics[width=\linewidth]{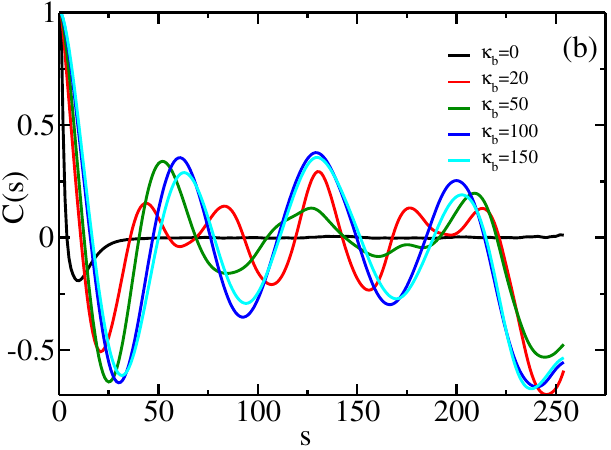}
\caption{The bond correlation $C(s)$ along the chain as a function of the arc-length (s) for various bending rigidities shown in the plot for the chain length $N=64$  (a) and $N=256$ (b)  at $\Gamma_e=5$. } 
\label{Fig:bond_corelation_256}
\end{figure}


\subsection{Bond Correlation} The circular state of the PA can also be identified from the bond-bond correlations; this provides insight into the correlated buckling of a semi-flexible chain\cite{gordievskaya2019conformational}. The correlation can be computed from one of the ends of PA, defined as: 
\begin{equation}
   C(s) = \left<\Vec{\bm t}_i \cdot \Vec{\bm t}_j \right>,
    \label{Eq:cos}
\end{equation}
where ${\bm t}_i$ and ${\bm t}_j$ are the unit bond vectors at the $i$th and $j$th  bond locations, and  $s$ is contour length $s=|i-j|l_0$.  Figure\ref{Fig:bond_corelation_256} displays the correlation for various bending rigidities for the diblock PA. The bond-bond correlation $C(s)$ diminishes exponentially to zero with arc length ($s$) for smaller bending rigidity, indicating the PA backbone lacks any long-range orientation.  Beyond a certain range of bending rigidity, the sinusoidal nature of correlation among bonds emerges, and their amplitude decays slowly with $s$. Further,  an increase in the bending leads to a decrease in the frequency with the larger amplitude of the oscillation, denoting the scale of correlation grows.  This damped oscillatory correlation implies the presence of circular conformations\cite{anand2018structure}.  As expected, the PA acquires a larger curvature for stiffer chains and longer chains, as shown in Fig.\ref{Fig:bond_corelation_256}. The length scale associated with the correlation can be estimated from an exponentially decaying sinusoidal function\cite{anand2018structure}, which turns out to be nearly the same as obtained from the radial distribution function.   

\subsection{Radial Distribution function}
The next quantity is {\it radial density distribution function} of the monomers from the center of mass of the PA chain. The radial distribution can be computed from the expression \cite{doi:86,Bolhuis_AEP_2001}, 
\begin{equation}
    g_{cm}(r) = \frac{n_{cm}(r)}{4 \pi r^2  n_0 dr }
    \label{Eq:gr}
\end{equation}
where $n_0$ is the mean local density of monomers and  $n_{cm}(r)$ is the local density of the monomers in the spherical cell of width $dr$ at a distance $r$ from the center of mass of the PA.

\begin{figure}[t]
\includegraphics[width=\linewidth]{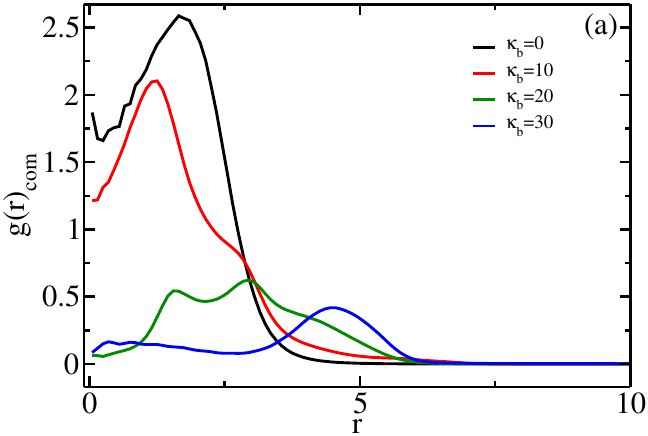}
\includegraphics[width=\linewidth]{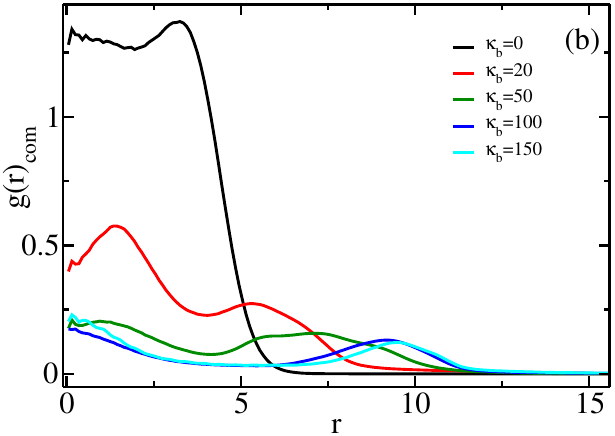}
\caption{ The radial distribution function w.r.t. center of mass of the chain for various bending strengths for diblock PA for chain lengths $N=64$ (a) and  $N=256$ (b) at $\Gamma_e=5$. }
\label{Fig:rdf_256}
\end{figure}

The two peaks in \ref{Fig:rdf_256} are due to the existence of extended and circular states simultaneously. The peak of the $g_{cm}(r)$ shifts from the center of the PA for the intermediate and large values of $\kappa_b$. Here, the radial density distribution function undergoes a qualitative change, suggesting that the monomers are distributed in a layer far from the center of the PA. This reveals the formation of a torus or circular structure, where the monomers are concentrated on a circle. The average radius of the torus/circular structures can be estimated from the peak's location in $g_{cm}(r)$.
For larger bending rigidities, the location of the peak in $g_{cm}$ shifts a larger distance from the center of mass, indicating the radius of the circular conformation gets bigger.

A simple quantitative argument for a diblock semi-flexible PA of length $L = 256$ can be given a roughly estimated radius of the circular/torus conformation, which is calculated as $r \approx  256/(2\pi N_t)=10$, where $N_t =4$ is the number of turns. Similarly, the six-fold torus phase exhibits a radius of 6 units for the smaller bending rigidities.
{This suggests for longer block lengths, the chain predominantly adopts the torus-like conformations with the different radii; however, they undergo the transition from one to another followed by various folded intermediate states; thus, they acquire many more meta-stable conformations than shorter block-length chains. }

\begin{figure}[t]
\includegraphics[width=\linewidth]{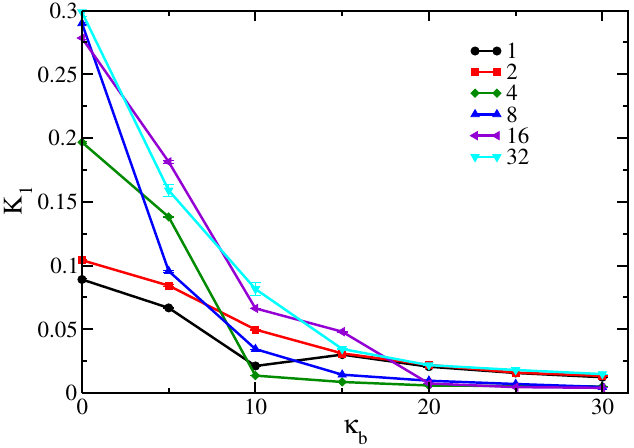}
\includegraphics[width=\linewidth]{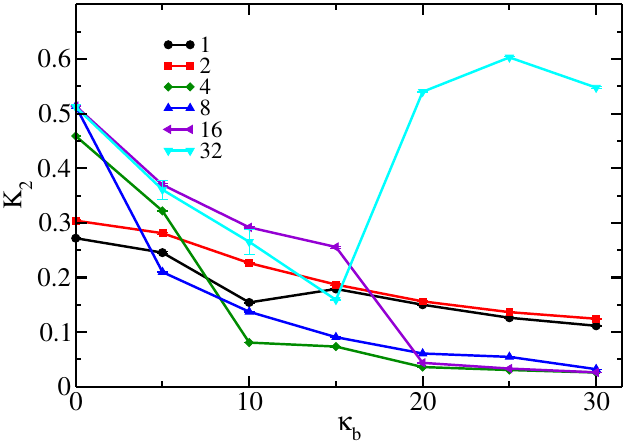}
\caption{ The average shape factors $K_1$ (a) and $K_2$ (b) for $\Gamma_e=5$ for chain length $N=64$.
}
\label{Fig:space_factor_k1_k2}
\end{figure}

\subsection{Shape Factor}

Figure \ref{Fig:space_factor_k1_k2} depicts the average values of shape factors $K_1$ and $K_2$ for various block lengths across varying bending rigidities. $K_1$ sharply decreases for $\kappa_b<10$ and for larger block lengths of $32,16$, $8$, and $4$. Further, an increase in the bending leads to monotonic descents for $\kappa_b>10$. For large values of $\kappa_b$, it acquires a minimal number $K_1< 0.05$. For the smaller block lengths $1$ and $2$, $K_1$ acquires smaller values, indicating the chain  acquires the elongated shape conformations, which further becomes extended for larger bending parameters as $K_1$ and $K_2$ diminishes.\\

In the case of diblock PA ($L_B=32$), as the bending rigidity increases, $K_2$ descends from 0.5 to 0.2, indicating the extension of the globular state and the possible coexistence of mixed conformations. A further increase in bending rigidity leads to a sudden rise in $K_2$, signifying the transition to circular conformations at $k_b\approx 20$. But for block lengths $16$, $8$, and $4$,  $K_2$ predominantly exhibits a decreasing trend with increasing bending rigidity, pointing to a transition from globule to hairpin-like conformation and 
for block length $2$ and $1$, the transition is from coil-like shape to elongated rod-like shape.

We present the normalized probability distribution of the shape factor $K_2$ for the diblock PA for various $\Gamma_e$ values. The figure indicates the distribution function has a single peak for smaller $\Gamma_e = 0.5$. For the larger $\Gamma_e \geq 2$, distribution becomes bimodal with $\kappa_b$. The bimodal distribution is the signature of the transition, indicating the coexistence of the two different structures.  The cumulative sum of the normalized probability distribution over specific ranges provides the most probable conformation.



\begin{figure}[ht]
\includegraphics[width=0.49\linewidth]{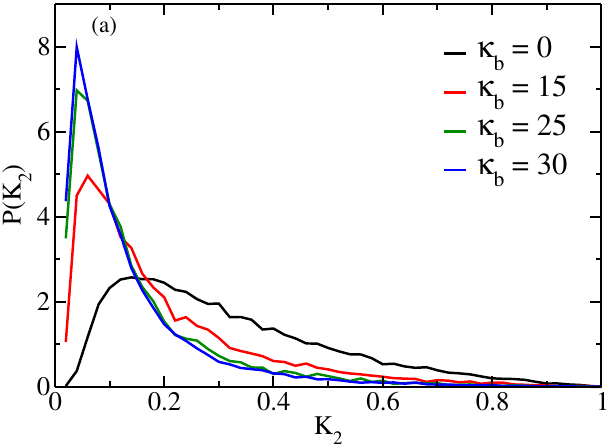}
\includegraphics[width=0.49\linewidth]{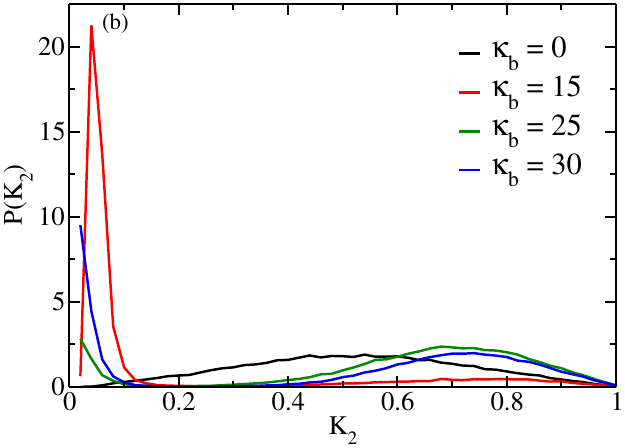}
\includegraphics[width=0.49\linewidth]{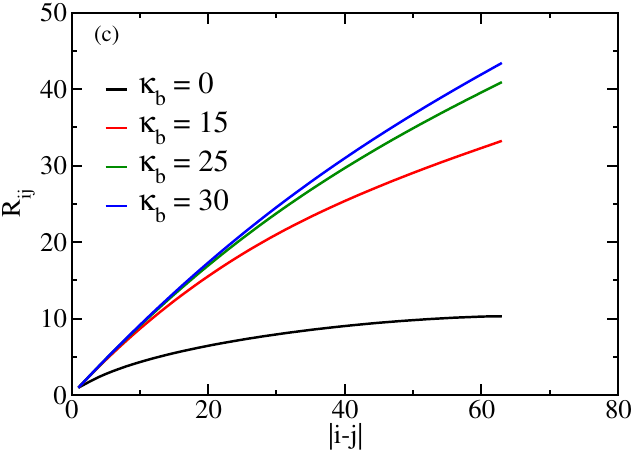}
\includegraphics[width=0.49\linewidth]{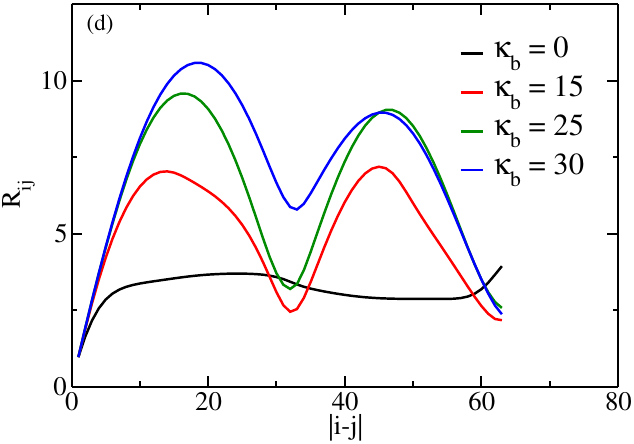}
\caption{The normalized probability distribution of the shape factor ($K_2$) and $R_{ij}$ for the case of block length $L_B=2$ (a,b) and 32 (c,d) at different bending rigidities at $\Gamma_e=5$. 
\label{Fig:rij}}
\end{figure}

The $R_{ij}$ profile of the PA continually changes on the variation of the $\Gamma_e$ and $\kappa_b$ from the single peak distribution to the double peak distribution. This ascertains a folded and circular hairpin-like state due to the strong attraction among oppositely charged blocks in the PA. The profile for different $\Gamma_e$ values is displayed in Fig.~\ref{Fig:rij}.

\begin{figure}[th]
\includegraphics[width=0.49\linewidth]{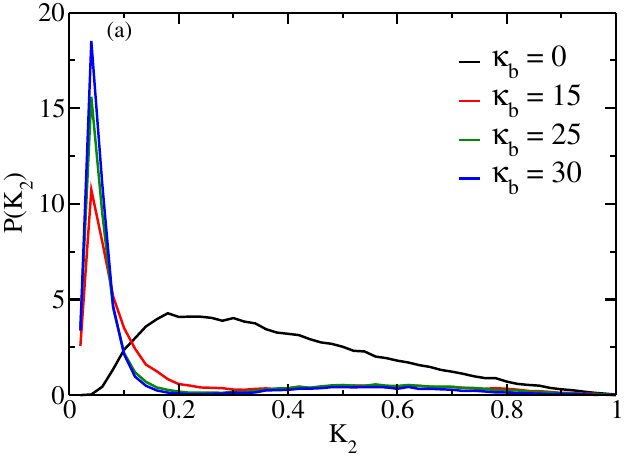}
\includegraphics[width=0.49\linewidth]{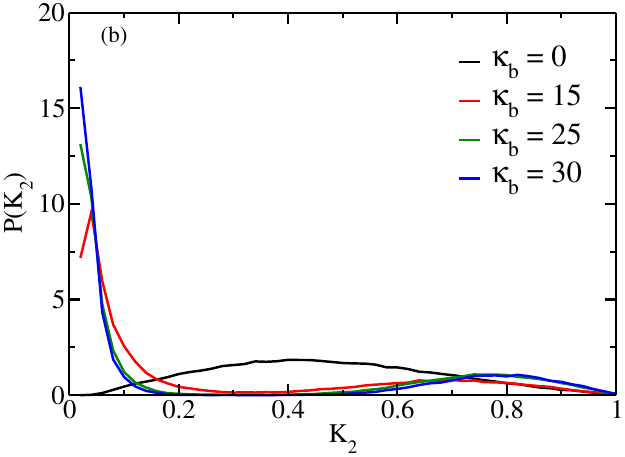}
\includegraphics[width=0.49\linewidth]{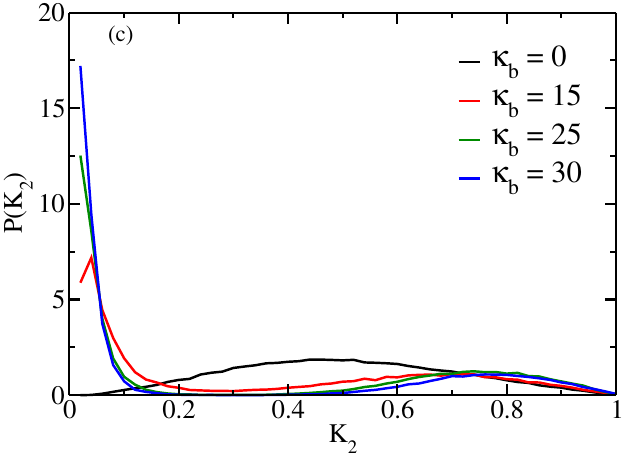}
\includegraphics[width=0.49\linewidth]{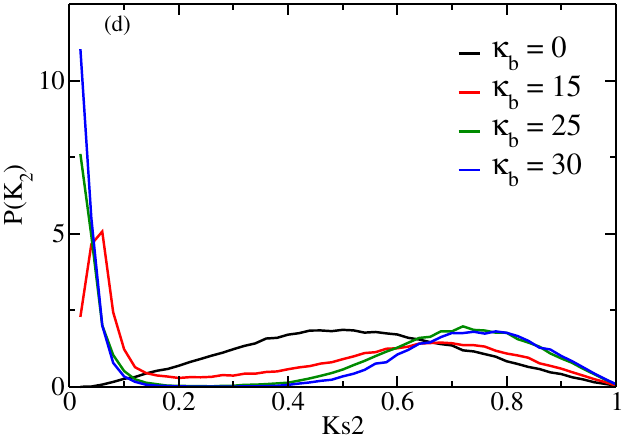}
\caption{The normalized probability distribution of the shape factor $K_2$ for the case of diblock PA at $\Gamma_e=0.5$ (a), $\Gamma_e=2$ (b), $\Gamma_e=3$ (c), and $\Gamma_e=4$ (d). 
\label{Fig:space_dist_lb}}
\end{figure}


\begin{figure}[t]
\includegraphics[width=0.49\linewidth]{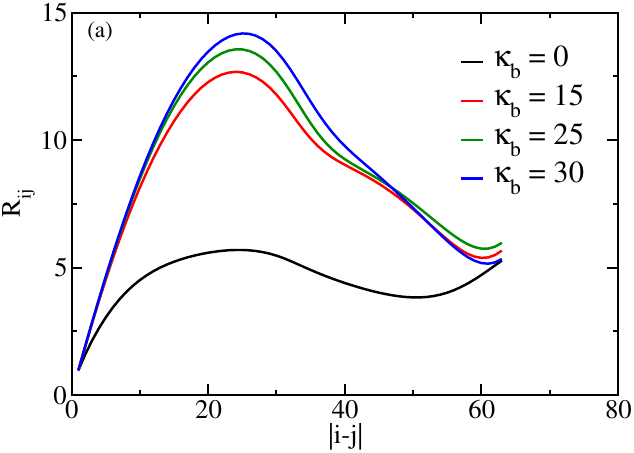}
\includegraphics[width=0.49\linewidth]{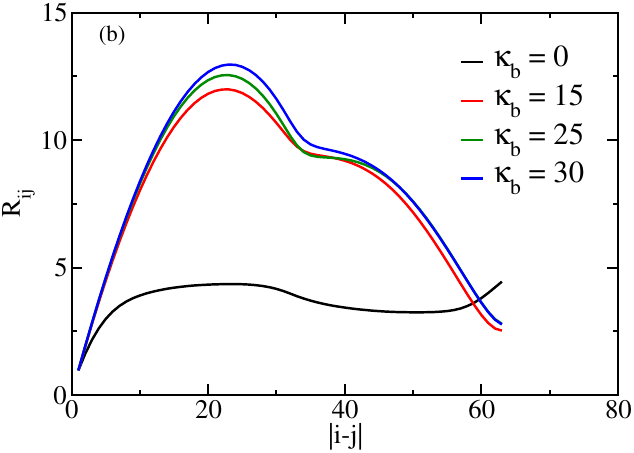}
\includegraphics[width=0.49\linewidth]{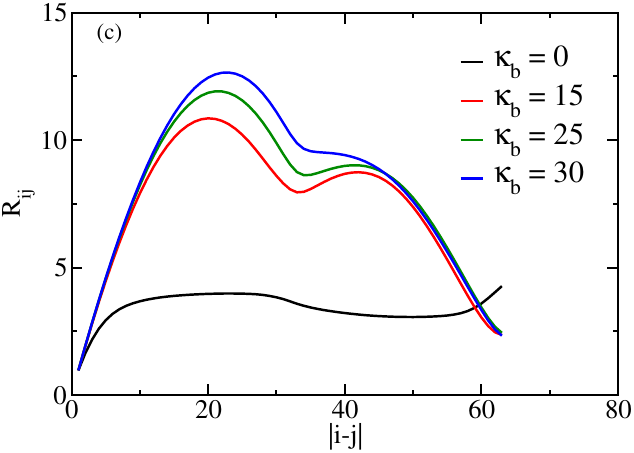}
\includegraphics[width=0.49\linewidth]{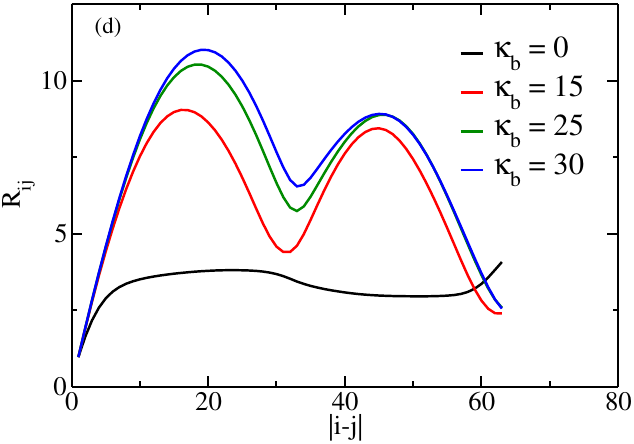}
\caption{The $R_{ij}$ profile for the case of diblock PA at $\Gamma_e=0.5$ (a), $\Gamma_e=2$ (b), $\Gamma_e=3$ (c), and $\Gamma_e=4$ (d). 
\label{Fig:rij_lb}}
\end{figure}

 \clearpage 

%